\documentclass[preprint,notoc]{JHEP3}
\usepackage{epsfig}

\def\eslt{E_T^{\rm miss}}
\def\to{\rightarrow}

\def\bi{\begin{itemize}}
\def\ei{\end{itemize}}
\def\te{\tilde e}

\def\tu{\tilde u}

\def\tb{\tilde b}
\def\tf{\tilde f}

\def\tst{\tilde t}
\def\ttau{\tilde \tau}
\def\tmu{\tilde \mu}
\def\tg{\tilde g}

\def\tq{\tilde q}

\def\tw{\widetilde W}
\def\tz{\widetilde Z}
\def\alt{\stackrel{<}{\sim}}
\def\agt{\stackrel{>}{\sim}}
\def\be{\begin{equation}}  
\def\ee{\end{equation}}  

\title{Collider and Dark Matter Searches in Models with \\
Mixed Modulus-Anomaly Mediated SUSY Breaking}

\author{Howard Baer$^a$, Eun-Kyung Park$^a$, Xerxes Tata$^b$ 
and Ting T. Wang$^a$\\
$^a$Department of Physics, Florida State University Tallahassee, 
FL 32306, USA\\
$^b$Department of Physics and Astronomy, University of Hawaii,
Honolulu, HI 96822, USA\\
E-mail: \email{baer@hep.fsu.edu},\email{epark@hep.fsu.edu},
\email{tata@phys.hawaii.edu},\email{tingwang@hep.fsu.edu} }

\preprint{\vbox{\hbox{FSU-HEP-060426, UH-511-1087-06}}} 

\abstract{ We investigate the phenomenology of supersymmetric models
where moduli fields and the Weyl anomaly make comparable contributions
to SUSY breaking effects in the observable sector of fields. This mixed
modulus-anomaly mediated supersymmetry breaking (MM-AMSB) scenario is
inspired by models of string compactification with fluxes, which have
been shown to yield a de Sitter vacuum (as in the recent construction by
Kachru {\it et al}).  
The phenomenology depends on the so-called modular weights which, in
turn, depend on the location of various fields in the extra dimensions.
We find that the model with zero modular weights gives 
mass spectra characterized by very light top squarks and/or tau
sleptons, or where $M_1\sim -M_2$ so that the bino and wino are
approximately degenerate.  The top squark mass can be in the range required by
successful electroweak baryogenesis.  The measured relic density of cold
dark matter can be obtained
via top squark co-annihilation at low $\tan\beta$, tau
slepton co-annihilation at large $\tan\beta$ or via bino-wino
coannihilation. Then, we 
typically find low rates for direct and indirect detection of neutralino
dark matter.  However, essentially all the WMAP-allowed parameter space
can be probed by experiments at the CERN LHC, while significant portions
may also be explored at an $e^+e^-$ collider with
$\sqrt{s}=0.5$--1~TeV. 
We also investigate a case with non-zero modular
weights. In this case, co-annihilation, $A$-funnel annihilation and bulk
annihilation of neutralinos are all allowed. Results for future
colliders are qualitatively similar, but prospects for indirect
dark matter searches via gamma rays and anti-particles are somewhat better. }


\begin{document}

\section{Introduction}
\label{sec:intro}

One of the main goals of string phenomenology is to connect
string theory to observable phenomena at colliding beam, or other,
experiments. This difficult enterprise may be made tractable by merging of
top-down and bottom-up approaches of connecting weak scale phenomena
to superstring theory valid at and above the string scale.
The many theoretical and phenomenological advantages of 
weak scale supersymmetry provide a target for what string theory must yield
at energy scales of order $\sim 1$ TeV. Alternatively, the discovery of 
weak scale supersymmetry and tabulation of the superparticle properties
could shed important light on the nature of physics at the string scale. 

A significant hurdle to the implementation of the string phenomenology program
is the existence of many flat directions in the space of
scalar fields (the moduli), since many physically relevant couplings are
determined by the ground state values of these moduli. Determining these
requires that the flat directions be lifted and stabilized, at least in a
controlled approximation, so that the ground state whose properties
determine low energy physics can be reliably determined. This has been
facilitated by a new class of compactifications, where the extra spatial
dimensions are curled up to small sizes with fluxes of additional fields
turned on along these extra dimensions. The presence of these fluxes
leads to calculable minima in the potential of the moduli, and
represents a starting point of the program for discovering a string ground
state that will lead to the (supersymmetric) Standard Model at low
energies, and which is consistent with constraints from cosmology. By
the latter, we refer to upper bounds on the energy density of moduli from the
age of the universe, to constraints on their decays after
nucleosynthesis, and to the observed acceleration of supernovae at high
red shifts which implies a de Sitter universe. 

Recently, Kachru, Kallosh, Linde and Trivedi (KKLT)\cite{kklt}  provided
a concrete model, based on type-IIB superstrings, of compactification 
with fluxes to a Calabi-Yau orientifold.
The background fluxes (non-zero vacuum expectation values of various
field strengths) allow one to stabilize the complex structure moduli
(that determine the shape of the compactified manifold) and the dilaton
field, and remove these fields from the low energy theory because they
are heavy.  The remaining size modulus is stabilized by a
non-perturbative mechanism such as gaugino condensation on a $D7$
brane. At this point, the vacuum of the theory is of anti-de Sitter
(AdS) type, in contradiction to cosmological observations. The addition
of a non-supersymmetric anti-brane ($\overline{D3}$) breaks the $N=1$
supersymmetry and lifts the vacuum minimum to zero or positive values,
yielding a de Sitter universe as required by cosmological measurements
referred to above. Due to the warping induced by the fluxes, the
addition of the $\overline{D3}$ brane also breaks supersymmetry by a
very small amount.  The resulting low energy theory thus has no unwanted
light moduli, has a broken supersymmetry, and
a positive cosmological constant.  There is, however, still a need for a
concrete implementation of the KKLT idea with an explicit Calabi-Yau
space and choice of fluxes that yields a ground state leading to the
observed low energy phenomenology.

Recently, several papers have analyzed the structure of the ensuing soft
supersymmetry breaking terms (SSB) in models based on the KKLT
proposal\cite{choi}.  A very interesting result that they find is that
these terms can receive comparable contributions from both
modulus (gravity) mediated contributions and the so-called anomaly mediated
contributions, their relative size depending on a phenomenological
parameter $\alpha$, defined in the next section. The anomaly-mediated
SUSY breaking (AMSB) contributions\cite{amsb} can be comparable because of
the mass hierarchy
\begin{equation}
m_{\rm moduli}\gg m_{3/2}\gg m_{\rm SUSY} ,
\label{eq:hierarchy}
\end{equation}
that develops. This hierarchy automatically
alleviates  phenomenological problems from late
decaying moduli and gravitinos that could disrupt, for instance, 
the predictions of light element abundances from Big Bang nucleosynthesis.
Aspects of the phenomenology of these models have recently been explored
by several groups \cite{choi3,flm,eyy}.  
In the notation of Ref. \cite{flm}, which we adopt in this paper,
in the limit $|\alpha| \to 0$ one obtains SSB terms 
which are pure AMSB with attendant tachyonic slepton squared
masses.\footnote{We warn the reader that $\alpha$ defined in
  Ref. \cite{choi3} differs from the definition in Ref. \cite{flm} that we
  use here by $\alpha_{\rm Ref.\cite{choi3}}=
    \frac{16\pi^2}{\ln(M_P/m_{3\over 2})}{1\over \alpha_{\rm our}}$.}
In the limit of large $|\alpha|$, one obtains dominantly modulus
mediation (MM), possibly
with nearly universal soft terms. For intermediate values of $|\alpha|$
of most interest to us, the problem of tachyonic sleptons is
absent, and the phenomenology is the most novel.\footnote{Of course, the
hallmark feature of AMSB models, that the induced SSB parameters
are scale invariant, no longer obtains.} Following
Ref. \cite{choi3}, we will refer to this scenario as the mixed
modulus-anomaly mediated SUSY breaking (MM-AMSB) model. 

Regardless of its theoretical motivation and stringy underpinnings, an
examination of the phenomenology of the MM-AMSB framework is interesting
in its own right. It represents a particular fusion of two well-studied
models
and can be regarded as a different, theoretically consistent and
phenomenologically viable framework for the exploration of supersymmetry
phenomenology. As we will see in the following, the SSB parameters depend
on the so-called matter and gauge field 
modular weights that characterize the location of
these fields in the extra spatial dimensions. In the absence of any
specific string model compactification to select out a particular
vacuum, we treat these as phenomenological numbers, different choices
of which lead to quite different characteristics of the sparticle spectrum,
and hence different SUSY phenomenology.

Several groups have begun the exploration of the phenomenology of
MM-AMSB scenario \cite{choi3,flm,eyy}. Of particular importance in this
regard is the nature of the lightest supersymmetric particle (LSP),
which we will take to be the neutralino, since by (\ref{eq:hierarchy}),
the gravitino is heavier than MSSM sparticles. First, this affects
the topology of SUSY events at colliders: for instance, a higgsino-like
neutralino LSP will couple more strongly to the third generation,
thereby increasing the $b$-jet\cite{mmt} and $\tau$-lepton multiplicity
in SUSY events\cite{prl}.  Second, the mass and composition of the LSP
sensitively affect its annihilation rate in the early universe, and
hence also the expected thermal dark matter LSP relic density that has
now been determined at better than the 10\% level by a study of the
cosmic microwave background to be \cite{wmap},
\begin{equation}
\Omega_{CDM}h^2=0.111^{+0.006}_{-0.01}\;,
\label{eq:wmap}
\end{equation}
where we quote the value obtained by the WMAP collaboration by combining
their three year result combined with the Sloan Digital Sky Survey
data.\footnote{The central value we use is almost unchanged from their
earlier result based on the analysis of just the first year WMAP
data. Our conclusions are insensitive to the precise value that we use.}
This measurement serves as a stringent constraint on any model with a
stable weakly interacting massive particle such as the neutralino in
$R$-parity conserving SUSY models\cite{wmapcon}. Third, the mass and
composition of the neutralino sensitively determine the prospects for
its detection in direct and indirect searches for dark matter
\cite{BBKOjcap}.
The character of the neutralino LSP varies widely depending on the
parameter $\alpha$, and also on the location of the gauge fields in the
extra dimensions. Finally, the authors of Ref. \cite{lnr} argue that,
even for heavy top squarks, for
some choices of MM-AMSB model parameters
the value of
$m_{H_u}^2(M_{\rm GUT})$ is largely cancelled by its renormalization
between $Q=M_{\rm GUT}$ and $Q=M_Z$; as a result,
these models may have less fine-tuning relative to other
frameworks.


The purpose of this paper is to map out the SUSY reach within the
MM-AMSB model framework in regions of model parameter space where
indirect constraints from rare $B$ and $B_s$ decays, from $(g-2)_\mu$, from
the DM relic density, and from direct sparticle and Higgs
boson searches are respected. Toward this end, we first map out the
parameter space regions consistent with the WMAP constraint
(\ref{eq:wmap}): agreement with (\ref{eq:wmap}) occurs due to a variety
of mechanisms, depending on where we are in parameter space. We then
delineate the SUSY reach of the CERN LHC and a $\sqrt{s}=0.5-1$ TeV
linear $e^+e^-$ collider in these regions. We also comment on particular
characteristics of the SUSY collider signatures for selected model
scenarios, and remark on the prospects for direct and indirect detection
of the neutralino LSP within this framework.  The remainder of this
paper is organized as follows.  We briefly review the KKLT construction,
and highlight the characteristics of the SSB parameters within the
MM-AMSB framework in Sec.~\ref{sec:model}.  In Sec. \ref{sec:zmw}, we
examine the mass spectrum of the model for the case of zero modular
weights for matter supermultiplets.
We find a rather large magnitude for the $\mu$ parameter, so that the
LSP is dominantly bino-like.\footnote{In this respect, our result
differs from that in Ref. \cite{flm}. We have discussed this with
Y. Mambrini who concurs with us that the discrepancy occurs because of
an error in the sign convention for the $A$-parameter used in
Ref. \cite{flm}. Once this is fixed, our results are in qualitative
agreement.}  The next-to-lightest supersymmetric particle (NLSP) is
found to be either a top squark or a tau slepton. In this case, WMAP
allowed regions are obtained where top squark or tau slepton
co-annihilation effects act to suitably reduce the neutralino relic
density.  The WMAP allowed regions give rather low rates for direct and
indirect detection of neutralino dark matter.  We also estimate the
reach of the CERN LHC and also a $\sqrt{s}=0.5-1$ TeV International
Linear Collider (ILC).  In our calculations, almost all the WMAP allowed
region of parameter space should be accessible to LHC searches.  In
Sec. \ref{sec:nzmw}, we examine a case with non-zero modular weights
(NZMW) for which the top squark can be more massive, so stop
co-annihilation can no longer occur. In this case, we show that $A$ and
$H$-funnel annihilation may be possible.
Finally, in Sec. \ref{sec:conclude}, we summarize our results
and present some conclusions.

\section{MM-AMSB model and soft SUSY breaking parameters}
\label{sec:model}

\subsection{The KKLT construction}
The KKLT construction\cite{kklt} realizes metastable de Sitter vacua with all
moduli stabilized. It breaks supersymmetry in a controlled way. In
this construction, one first introduces nonzero fluxes in the Type IIB
string theory compactified on a Calabi-Yau manifold. Due to the
nonzero fluxes,  the complex structure moduli and the dilaton are
completely fixed. However, the size modulus $\hat{T}$ remains a flat direction and
is still not fixed. To fix this, KKLT invoke
non-perturbative effects, such
as gaugino condensation on $D7$ branes. 
At this stage,
all moduli are fixed, but one ends up with supersymmetric vacua and 
negative vacuum energy.  
The final
step in the construction is to include anti $D$-branes yielding the desired
de-Sitter vacua (with positive vacuum energy)
and breaking supersymmetry.  Because of the presence
of branes   
and fluxes, the models have generically warped compactifications. Due
to the warping, the addition of the anti $D$-brane breaks supersymmetry
by a very small amount. 

Since the shape moduli are heavy enough to be integrated out, the low
energy theory can be regarded as a broken supergravity theory of the size
modulus $\hat{T}$. The K\"ahler potential for $\hat{T}$ is the no-scale
type, and an exponential superpotential for it is induced by
non-perturbative effects. 
Analysis\cite{choi} shows the modulus-mediated
contribution to the SSB parameters is roughly
\begin{equation}
\frac{F^T}{T+T^*}=  \mathcal{O}\left(\frac{m_{3/2}}{\ln
  (M_{P}/m_{3/2})}\right) \sim \frac{m_{3/2}}{4\pi^2}\;,
\label{eq:mm}
\end{equation}
where $T$ denotes
the scalar component and $F^T$ the auxiliary component of the size
modulus.  To obtain the last equality, we assume $m_{3/2}\sim 1$~TeV.
Because of the additional suppression by the large logarithm, we see
that the tree-level modulus-mediated contributions to MSSM SSB
parameters can be comparable to the corresponding AMSB contribution,
whose scale is given by, 
\begin{equation}
m_{\rm soft}\sim \frac{m_{3/2}}{8\pi^2} .
\label{eq:amsb}
\end{equation}
The original KKLT proposal predicted the relative size of the modulus
and AMSB contributions. It is, however, possible to generalize this picture
so that the ratio of these contributions is determined by a
phenomenological parameter $\alpha$ (that can have either sign) as we
have already mentioned \cite{choi3,flm}. 
We note that mixed modulus-anomaly mediated contributions to SSB
parameters may also be seen in some of the ``benchmark models'' of 
Ref. \cite{Kane:2002qp}, where references to the literature for their
theoretical basis may be found.

The gauge kinetic functions and the K\"ahler potential for the visible matter
superfields ${\hat{Q}}_i$
depend on their location in the extra dimensions. The gauge kinetic
function is given by, 
\begin{equation}
f_a={\hat{T}}^{\ell_a},
\end{equation}
where $a$ labels the gauge group and the power $\ell_a= 1 \ (0)$ for
gauge fields on $D7$ ($D3$) branes.
The K\"ahler potential for the matter fields is
\begin{equation}
K=\sum_i \frac{1}{(\hat{T}+\hat{T}^*)^{n_i}} {\hat{Q}}_i^*{\hat{Q}}_i ,
\end{equation}
with the modular weights $n_i=0 \ (1)$ for matter fields located
on $D7$ ($D3$) branes; fractional values $n_i=1/2$ are also possible for
matter living at brane intersections \cite{choi3}.  From the gauge
kinetic functions, the K\"ahler potential and the superpotential, one
can calculate visible field SSB parameters that are required for SUSY
phenomenology.

\subsection{MM-AMSB model parameter space and soft terms}

The MM-AMSB model is completely specified by the parameter set,
\begin{equation}
\ m_{3/2},\alpha ,\ \tan\beta ,\ sign(\mu ),\ n_i,\ \ell_a.
\end{equation}
The mass scale for the SSB parameters 
is dictated by $m_{3/2}$, where $m_{3/2}$ is the gravitino mass. The parameter
$\alpha$ gives the relative contributions of 
anomaly mediation and gravity mediation to the soft terms, 
 $n_i$ are the modular weights of
the visible sector matter fields, and $\ell_a$ appears in the
gauge kinetic function. 
We see from (\ref{eq:mm}) and (\ref{eq:amsb}) for $|\alpha|=
\mathcal{O}(1)$ that the scale for the
SSB parameters of the visible sector is  $\sim {m_{3/2}\over {8\pi^2}}$.

More specifically, these SSB gaugino mass parameters, trilinear SSB
parameters and sfermion mass parameters, all renormalized just below the
unification scale (which we take to be $Q=M_{\rm GUT}$), are
respectively given by,
\begin{eqnarray}
M_a&=& M_s\left( \ell_a \alpha +b_a g_a^2\right),\label{eq:M}\\
A_{ijk}&=& M_s \left( -a_{ijk}\alpha +\gamma_i +\gamma_j +\gamma_k\right),
\label{eq:A}\\
m_i^2 &=& M_s^2\left( c_i\alpha^2 +4\alpha \xi_i -
\dot{\gamma}_i\right) ,\label{eq:m2}
\end{eqnarray}
where $$M_s\equiv\frac{m_{3/2}}{16\pi^2},$$
$b_a$ are the gauge $\beta$ function coefficients for gauge group $a$ and 
$g_a$ are the corresponding gauge couplings. The coefficients that
appear in (\ref{eq:M})--(\ref{eq:m2}) are given by, 
$$c_i =1-n_i,$$ $$a_{ijk}=3-n_i-n_j-n_k,$$ and
$$\xi_i=\sum_{j,k}a_{ijk}{y_{ijk}^2 \over 4} - \sum_a l_a g_a^2
C_2^a(f_i).$$ 
Finally, $y_{ijk}$ are the superpotential Yukawa couplings,
$C_2^a$ is the quadratic Casimir for the a$^{th}$ gauge group
corresponding to the representation to which the sfermion $\tf_i$ belongs,
$\gamma_i$ is the anomalous dimension and
$\dot{\gamma}_i =8\pi^2\frac{\partial\gamma_i}{\partial \log\mu}$.
Expressions for the last two quantities involving the 
anomalous dimensions are presented in
the Appendix of Ref. \cite{flm}, and will not be repeated here.
For brevity, we will sometimes use the notation that 
$A_t\equiv A_{Q_3H_u u_R}$, $A_b\equiv A_{Q_3H_d d_R}$ and
$A_\tau\equiv A_{L_3H_d e_R}$ .

Throughout our study, we set $\ell_a=1$, but examine the phenomenology for
various choices of modular weights, beginning with $n_i=0$ in the
next section.

\section{MM-AMSB Model with zero modular weights}
\label{sec:zmw}

\subsection{Soft SUSY breaking terms}

Following Ref. \cite{flm}, we first examine the MM-AMSB model with
modular weights $n_i=0$, and $\ell_a =1$. 
In this case, the contributions from modulus mediation are universal,
and for large $\alpha$ the mass pattern reduces to that of minimal
supergravity with $m_0=m_{1/2}=-A_0/3$. 
The values of
SSB parameters, renormalized at $Q=M_{\rm GUT}$, are plotted in
Fig. \ref{fig:soft_zmw} versus $\alpha$ for $m_{3/2}=11.5$ TeV,
$\tan\beta =10$ and $\mu >0$. We take $m_t=175$ GeV throughout this paper.
In frame {\it a}), we show the gaugino masses and $A_i$-parameters,
while in frame {\it b}) third generation sfermion and Higgs boson scalar
mass parameters are shown as $sign(m_i^2)\cdot \sqrt{|m_i^2|}$.  We see
from frame {\it a}) that for $\alpha =0$, the familiar pattern of AMSB
gaugino masses results: $M_1>M_2$, while $M_3<0$. As $\alpha$ increases,
all gaugino masses become positive. Moreover, because the differences
$M_1-M_2$ and $M_1-M_3$ are independent of $\alpha$, the gaugino masses
become essentially equal for large values of $|\alpha|$, as we had
anticipated above.\footnote{The differences between gaugino masses would
increase with $\alpha$ if the weights $\ell_a$ were dependent on $a$, but
this would mean that the gauge fields of the different factors of the
gauge group have different locations in the extra dimensions, precluding
the possibility of Grand Unification.} Since the GUT scale gaugino
masses do not depend on the modular weights for matter fields, their
behavior in frame {\it a}) also holds for models with non-zero modular
weights considered in the next section. 


For zero modular weights, the $A_i$ parameters would also be
universal if the AMSB contributions were small relative to the
modulus-mediated contributions, but instead they suffer some
splitting. 
For $|\alpha|\agt 2$, the universal modulus-mediated contributions 
dominate the AMSB contributions.
As we shall see, the large $A_t$ term acts
via RG running to suppress the scalar squared mass, $m_{\tst_R}^2$, so
that the top squark $\tst_1$ is frequently the next-to-lightest SUSY
particle (NLSP).

\FIGURE[htb]{
\epsfig{file=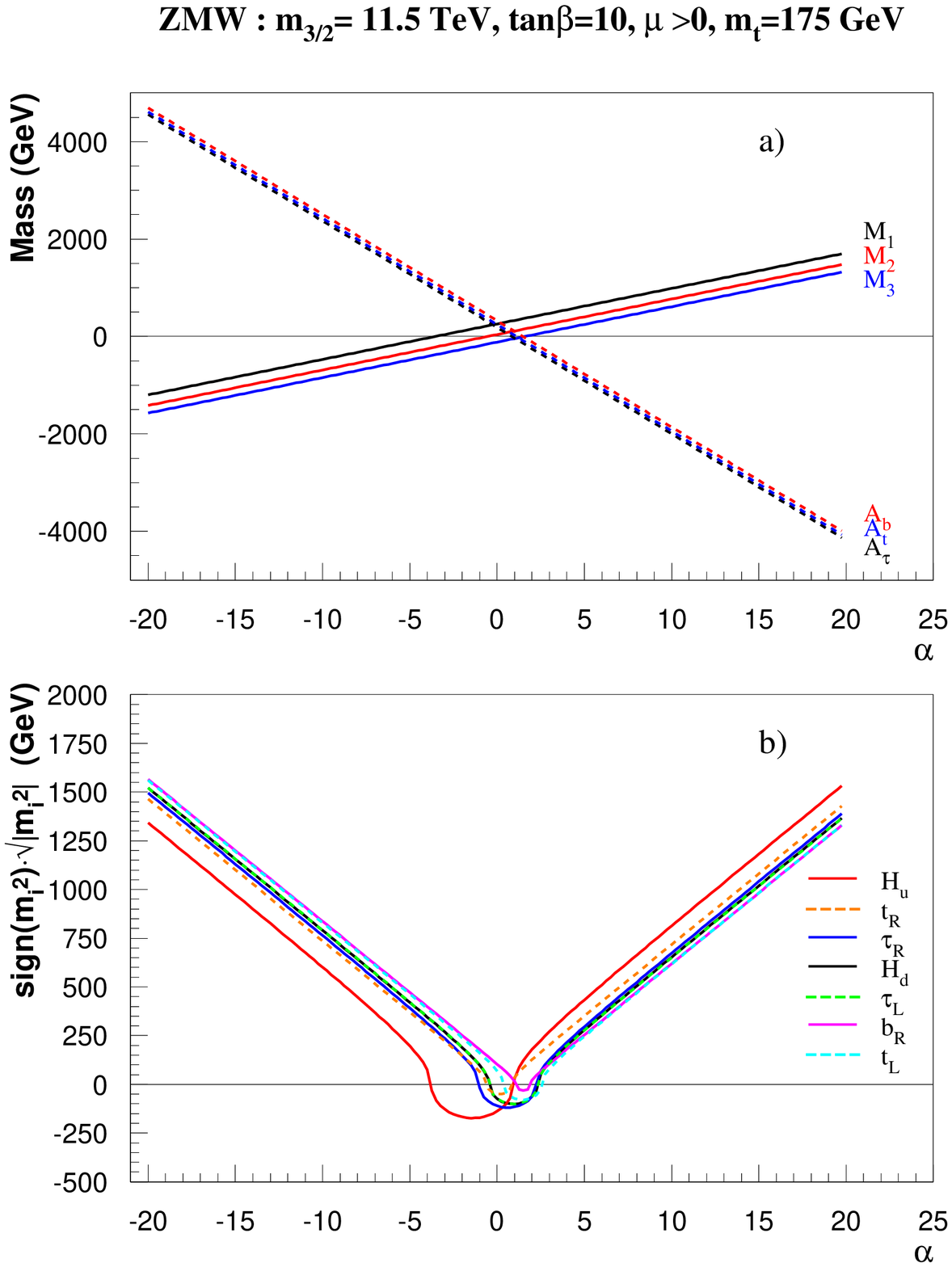,width=12cm} 
\caption{\label{fig:soft_zmw}
Various soft SUSY breaking parameters at the scale
$Q=M_{GUT}$ versus $\alpha$ for $n_i=0$, $\ell_a =1$, 
$m_{3/2}=11.5$ TeV, $\tan\beta =10$, $\mu >0$ and $m_t=175$ GeV.
In {\it a}), we show gaugino masses and $A$ terms, while in 
{\it b}) we show $sign(m_i^2)\cdot\sqrt{|m_i^2|}$ for third generation
scalar and
Higgs boson soft SUSY breaking masses.}}

The SSB masses for third generation and Higgs scalars are shown in 
Fig. \ref{fig:soft_zmw}{\it b}). For large values of $|\alpha|$, 
the $\alpha^2$ term in Eq.~(\ref{eq:m2}) dominates and we get the linear
dependence seen in the figure. As the magnitude of $\alpha$ reduces, the 
other terms grow in importance leading to the curvature in the
neighbourhood of $|\alpha|\alt 5$.  We also see that for
$\alpha$ close to zero, we reproduce the well-known tachyonic slepton 
squared masses of AMSB. 
This does not necessarily
mean that we are in the wrong vacuum because we may expect large
logarithmic corrections to the ``tree level potential with parameters
renormalized at the GUT scale''. Indeed after renormalization group
evolution to low scales, we see that over a portion of this range of
$\alpha$ we find acceptable spectra at the weak scale.
\footnote{In the AMSB framework, $t$-squark masses are also negative at
  the GUT scale, but evolve to positive values at low scales. Negative
soft masses for Higgs scalars have been considered previously in \cite{nuhm2},
while negative matter scalar soft masses have also been considered by Feng
{\it et al.}\cite{feng}.}
For moderate to large values of $|\alpha|$, the moduli-mediated
contributions dominate, and $m_{\tst_L}^2$ and $m_{\tb_R}^2$
($m_{\tst_R}^2$) are the smallest of the squark soft breaking terms at
the GUT scale for positive (negative) values of $\alpha$.

In Fig. \ref{fig:mi_evol}{\it a}), we illustrate the evolution of the
three gaugino masses from $Q=M_{GUT}$ to $Q=M_{\rm weak}$, for $\alpha =6$,
$m_{3/2}= 11.5$ TeV, $\tan\beta =10$ and $\mu >0$. While the soft terms
are ordered as $M_1>M_2>M_3$ at $M_{GUT}$, the RG evolution leads to the
familiar weak scale ordering, $M_3>M_2>M_1$, expected in models with
gaugino mass unification. As a result, unless $|\mu| \alt M_1({\rm
weak})$, the LSP is likely to be bino-like. Note, however, that the
ratios of weak scale gaugino masses is now $M_1:M_2:M_3 \sim 1:1.3:2.6$,
and differs sharply from $M_1:M_2:M_3 \sim 1:2:7$ in models with gaugino
mass unification at the GUT scale, or in gauge-mediated SUSY breaking
models. The presence of light gluinos in this framework will enhance the
reach of hadron colliders over electron-positron colliders. The most
striking feature of the figure is the well-understood phenomenon of
``mirage unification''\cite{eyy,choi3}, where from a weak scale
perspective, it appears that the gaugino masses unify at the
intermediate scale $Q\sim 10^{11}$ GeV; gauge couplings, however,
continue to unify at $Q= M_{\rm GUT}$.

\FIGURE[htb]{
\epsfig{file=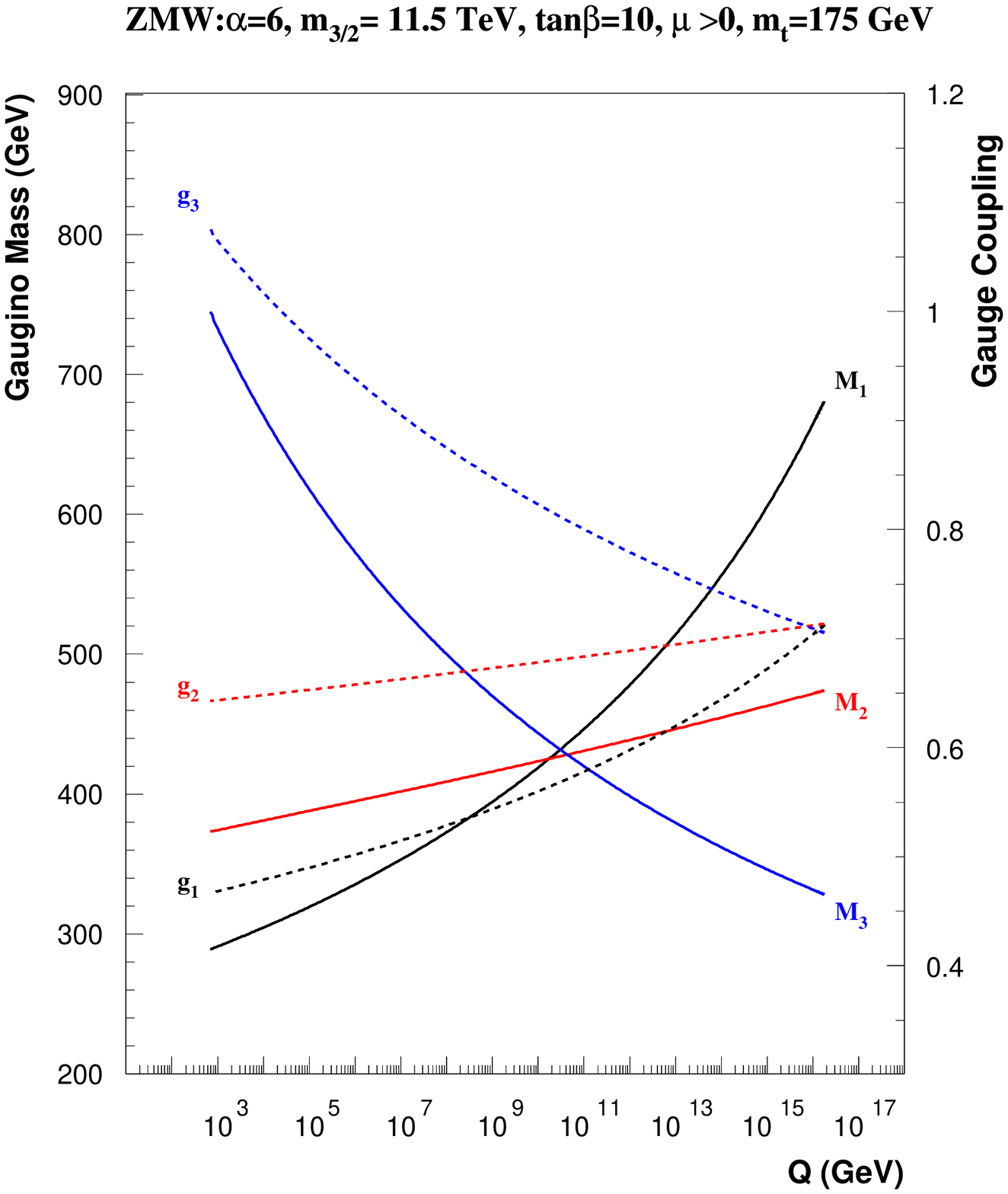,width=7cm} 
\epsfig{file=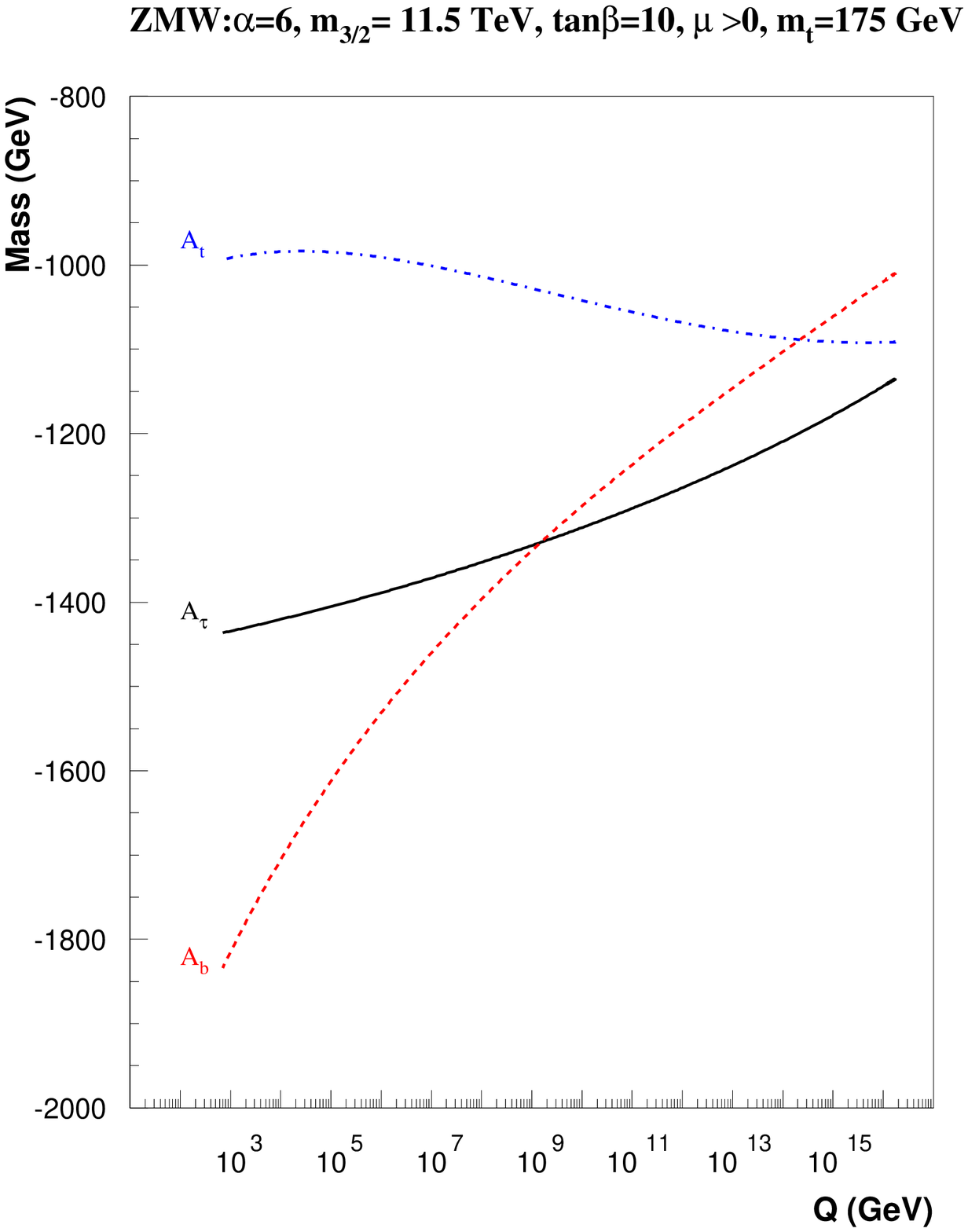,width=7cm} 
\caption{\label{fig:mi_evol}
Evolution of {\it a}) the gaugino masses $M_1$, $M_2$ and $M_3$, and
of {\it b}), the trilinear 
soft masses $A_t$, $A_b$ and $A_\tau$, from
$Q=M_{GUT}$ to $Q=M_{\rm weak}$ for the model with zero 
modular weights with $\alpha =6$,
$m_{3/2}=11.5$ TeV, $\tan\beta =10$, $\mu >0$ and $m_t=175$ GeV. Also
shown (right hand scale) is the corresponding evolution of the three gauge
couplings. }}

In Fig. \ref{fig:mi_evol}{\it b}), we show the evolution of the $A_i$
parameters versus $Q$ for the same parameter choice as in 
frame {\it a}). In this case, the evolution of $A_t$
is rather flat, while the magnitudes of $A_b$ and $A_\tau$
actually increase. 
The $A_i$ term RGEs read
\begin{eqnarray}
\frac{dA_t}{dt}&=&{2\over{16\pi^2}}\left(\sum_{i}c_ig_i^2M_i+
6f_t^2A_t+f_b^2A_b\right),\\
\frac{dA_b}{dt}&=&{2\over{16\pi^2}}\left(\sum_{i}c_i'g_i^2M_i+
6f_b^2A_b+f_t^2A_t+f_\tau^2A_\tau \right),\\
\frac{dA_\tau}{dt}&=&{2\over{16\pi^2}}\left(\sum_{i}c_i''g_i^2M_i+
3f_b^2A_b+4f_\tau^2A_\tau\right),\\
\end{eqnarray}
where $c_i=({13\over 15},3,{16\over 3})$, $c_i'=({7\over 15}, 3,
{16\over 3})$ and $c_i''=({9\over 5},3,0)$. On the right hand side, the
terms involving gauge couplings push the (already negative) $A_i$
parameters to more negative values, while the Yukawa terms push towards
less negative values. The large top quark Yukawa coupling $f_t$
counterbalances the gauge terms to yield a nearly flat running for
$A_t$, while the smaller $f_b$ and $f_\tau$ Yukawa couplings are not
sufficient to counterbalance the push of the gauge terms. For negative
values of $\alpha$, the gaugino mass parameters are negative, but
$A$-parameters start off positive, and the cancellation between the
gauge and Yukawa contributions to the evolution of $A$-parameters
persists. 

The weak scale values of the gaugino masses are shown in
Fig.~\ref{fig:weakgaug} for the same slice of parameter space as in
Fig.~\ref{fig:mi_evol}. The gaugino mass parameters are essentially
independent of matter modular weights (which enter only via sfermion
masses either via two-loop terms in the RGEs, or via sparticle
decoupling).  The striking feature of this figure is that for $\alpha
\sim 2.5$, the mirage unification now occurs essentially at the weak
scale. Close to this value of $\alpha$, $M_1({\rm weak})=M_2({\rm weak})$ so
that it is possible that mixed wino dark matter (MWDM), if allowed
within this framework by other constraints, is a viable DM candidate\cite{mwdm}. Likewise, for small and negative values of $\alpha$,
$M_1({\rm weak}) \sim -M_2({\rm weak})$, and bino-wino coannihilation (BWCA)
offers a viable possibility for DM in agreement with
Eq.~(\ref{eq:wmap}) \cite{bwca}. We should mention that, depending on
modular weights, different ranges of
$\alpha$ are excluded because of theoretical constraints: either
electroweak symmetry is not properly broken, or the LSP is a charged particle.
These ranges are shown for the model with zero modular weights on the
upper scale, and for a model with the choice $n_{H_u}=n_{H_d}=1$, $n_{\rm
  matter}=\frac{1}{2}$ considered in the next section. 
\FIGURE[htb]{
\epsfig{file=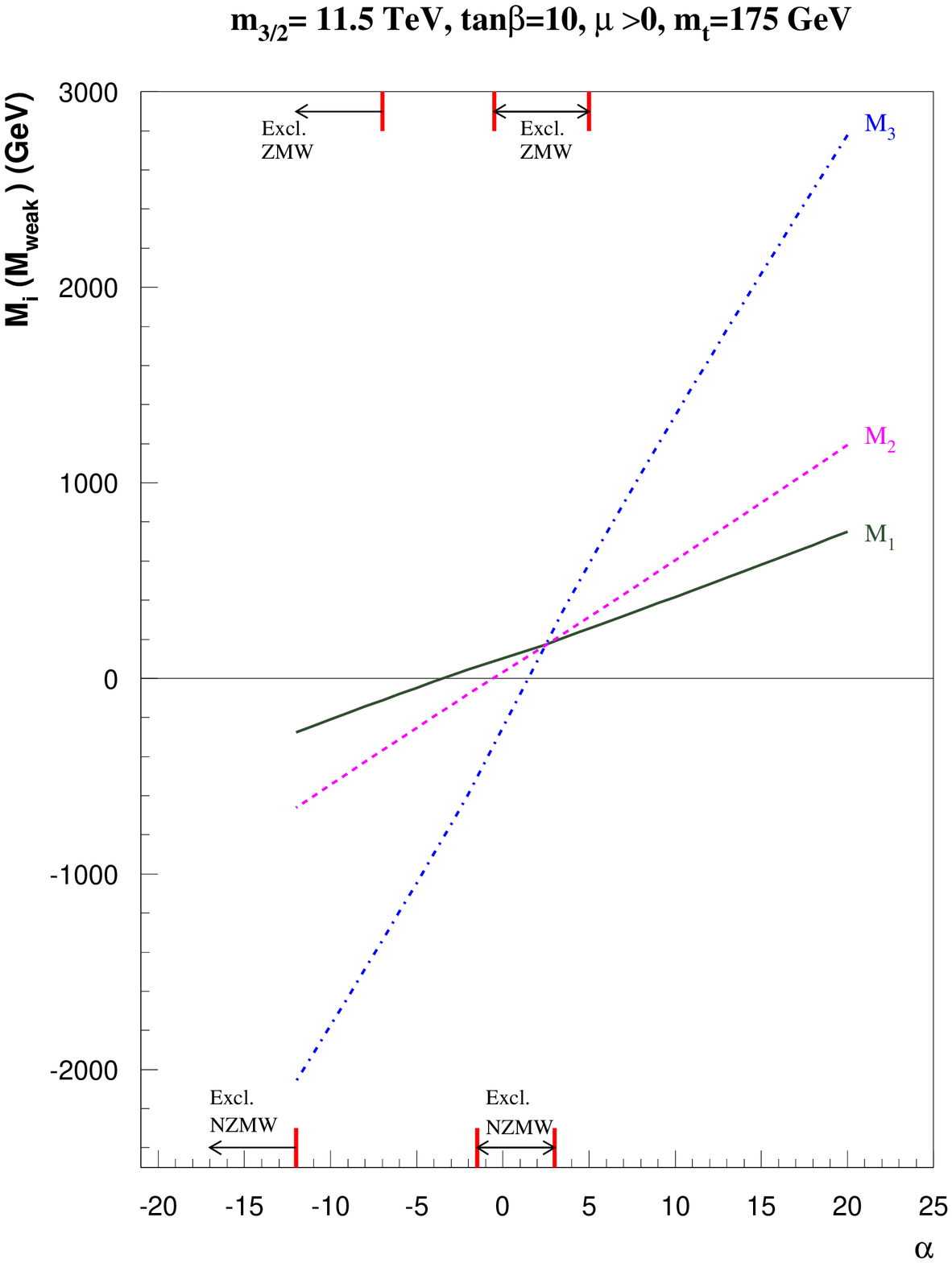,width=11cm}  
\caption{\label{fig:weakgaug} SSB gaugino mass parameters at the weak
  scale versus the MM-AMSB model $\alpha$ for $m_{3/2}=11.5$~TeV,
  $\tan\beta=10$, $\mu > 0$. For a value of $\alpha$ close to the mirage
  unification point, mixed wino dark matter may be possible, while for
  $\alpha \sim -1.75$, bino-wino coannihilation could lead to an
  acceptable dark matter relic density provided these values of $\alpha$
  are allowed by other constraints. The ranges of $\alpha$ excluded for
  the model with zero modular weights (ZMW) for matter fields are shown
  on the top, while the corresponding excluded ranges for the particular
  choice of non-zero modular weights (NZMW) $n_{H_u}=n_{H_d}=1$, $m_{\rm
  matter}=1/2$, is shown at the bottom. }}

Returning to the RGE of SSB parameters, in Fig. \ref{fig:sm_evol}{\it a}) 
we show the evolution of first generation scalar soft breaking
masses. Here, we see that the GUT scale mass ordering
$m_{e_R}^2>m_{e_L}^2>m_{d_R}^2>m_{u_R}^2>m_{u_L}^2$ (but with rather
small splittings) becomes essentially inverted at the weak scale, mainly
because of the large evolution of squark masses on account of SUSY QCD
interactions.  The mirage unification of scalar soft terms at $Q\sim
10^{11}$ GeV is also evident. In Fig. \ref{fig:sm_evol}{\it b}) we show
the evolution of third generation and Higgs boson soft scalar mass
parameters.  Since their evolution depends on Yukawa couplings, the
mirage unification no longer obtains: notice, however, that the $H_d$
and $\tb_R$ mass parameters (for which the Yukawa couplings are small)
do intersect close to the mirage unification scale in frame {\it a}).
We also see that $m_{\tst_R}^2$ evolves to much lower values than other
squark masses and, in this case, is not very different from the
corresponding {\it slepton} and wino mass parameters.
This effect, along 
with large mixing in the top squark mass matrix, leads to $m_{\tst_1}$
being the NLSP in the MM-AMSB model with zero modular weights and 
low $\tan\beta$.
\FIGURE[htb]{
\epsfig{file=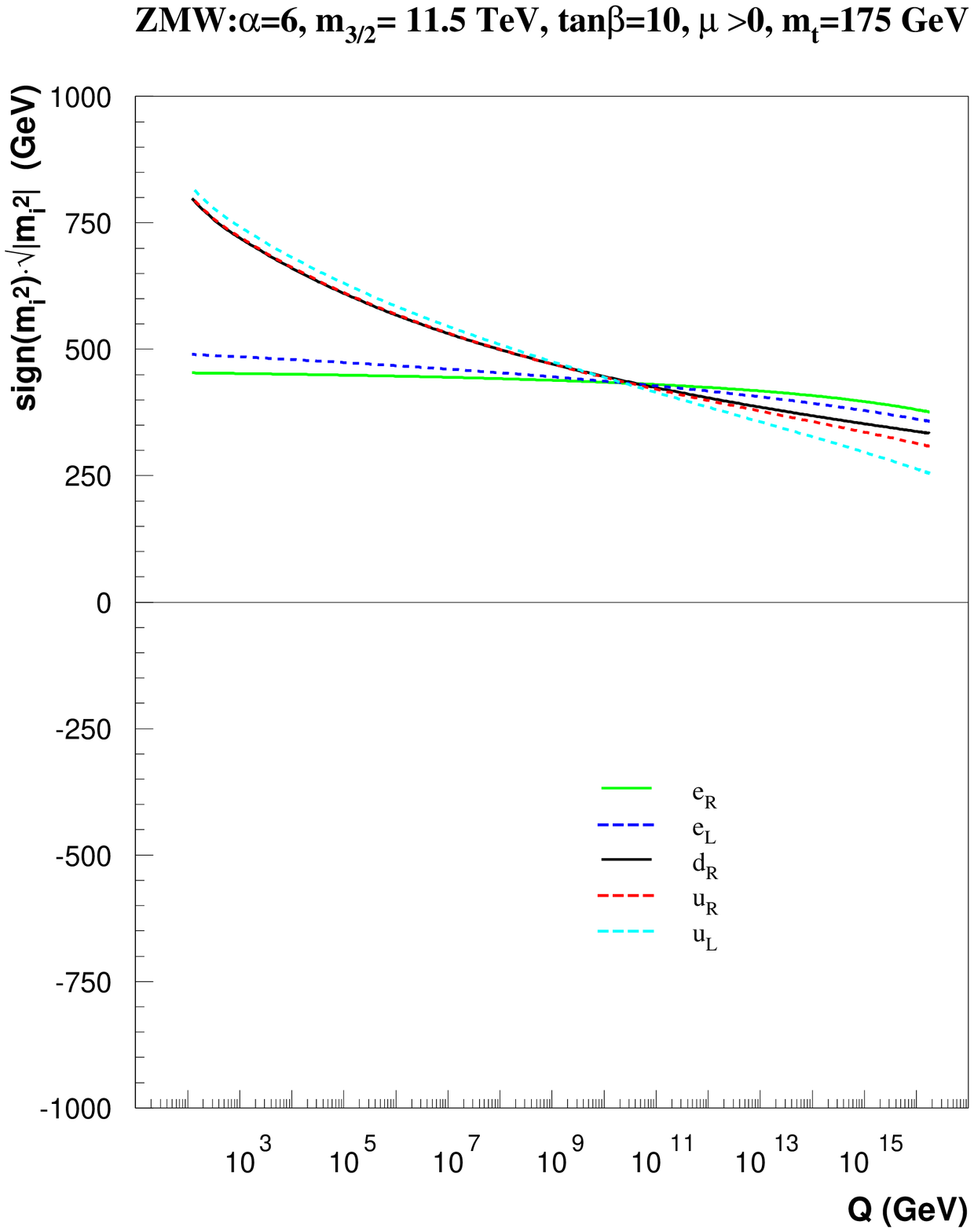,width=7cm} 
\epsfig{file=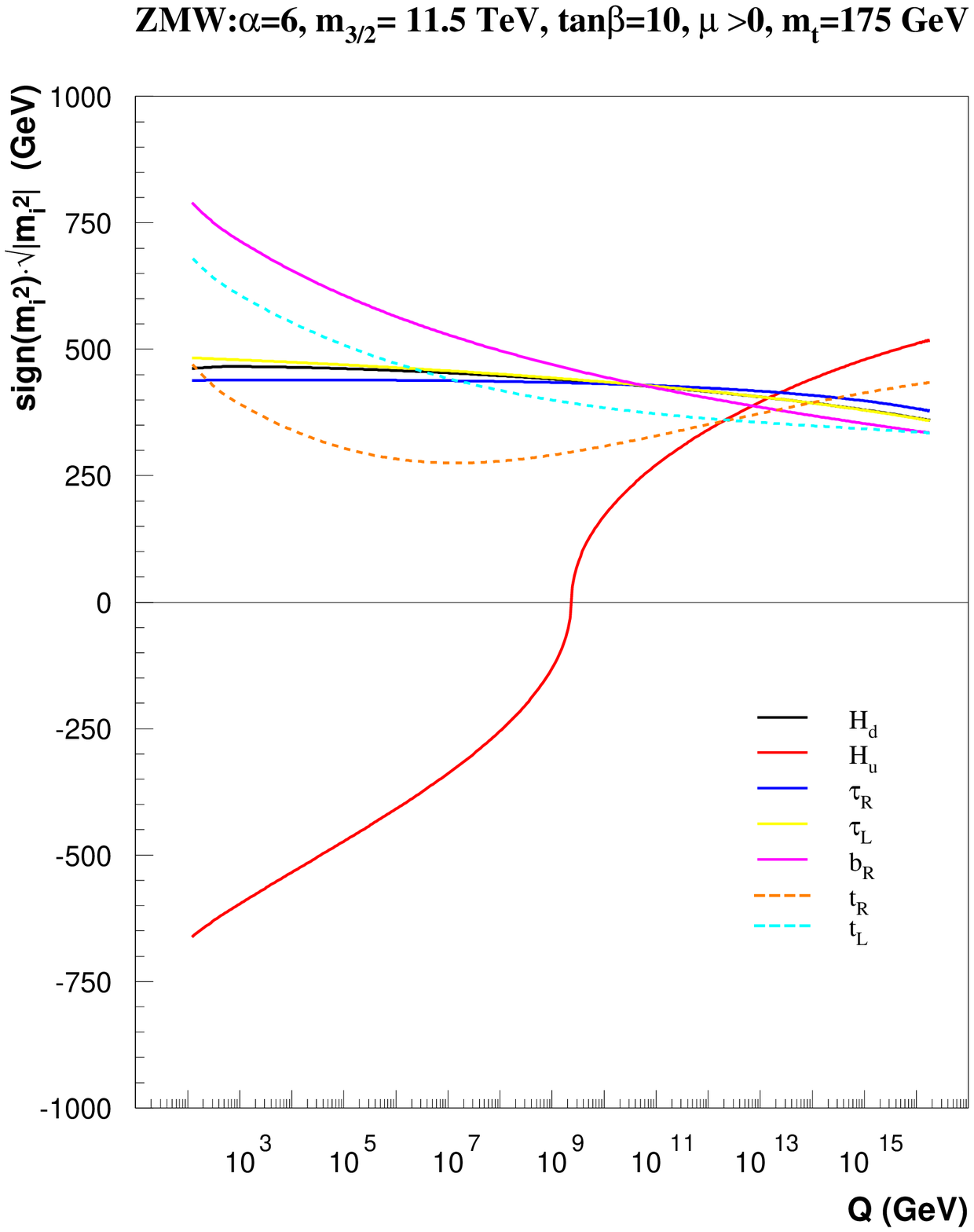,width=7cm} 
\caption{\label{fig:sm_evol}
Evolution of scalar soft masses $m_i^2$ for {\it a}) first generation
scalars and {\it b}) third generation and Higgs scalars from
$Q=M_{GUT}$ to $Q=M_{\rm weak}$ for $\alpha =6$,
$m_{3/2}= 11.5$ TeV, $\tan\beta =10$ and $\mu >0$, with $\ell_a=1$ and
modular weights set to zero. To show evolution to negative
squared masses,  
we actually plot $sign(m_i^2)\cdot\sqrt{|m_i^2|}$.}}

\subsection{Mass spectrum}

Once the soft SUSY breaking terms are stipulated at $Q=M_{GUT}$, then we
use Isajet v7.74\cite{isajet} 
to compute the corresponding weak scale mass spectrum.
In Fig. \ref{fig:mass_zmw_al}, we show sparticle masses and the weak
scale $\mu$ parameter vs. $\alpha$ for $m_{3/2}=11.5$ TeV, $\mu >0$ and
{\it a}) $\tan\beta =10$ and {\it b}) $\tan\beta =30$.  The red-shaded
region is excluded by lack of REWSB, and the blue-shaded region gives a
charged or colored ($\tst_1$ or $\ttau_1$) LSP, so that viable mass
spectra are only achieved for $\alpha \agt 5.5$ or $\alpha \alt -1.2$
The lines end at large
values of $|\alpha|$ because electroweak symmetry is not correctly
broken.  Since $|M_1|<M_2<M_3$ and $|\mu | \gg M_1$, the LSP $\tz_1$ is
bino-like, and the corresponding relic density is typically large and
beyond the WMAP bound. However, we see in frame {\it a}) that at low
$\alpha \sim 6$, $m_{\tst_1}\sim m_{\tz_1}$ so that top squark
co-annihilation is possible, which can act to reduce the relic density,
while for $\alpha$ just smaller than $-2$, the BWCA mechanism could be
operative as anticipated above.  For this choice of modular
weights and $\tan\beta$, MWDM is not possible, because $\tst_1$ becomes
the LSP for the required value of $\alpha$.  In the case of frame {\it
b}) with a larger value of $\tan\beta =30$, we find for low positive
$\alpha$ that $m_{\ttau_1}\sim m_{\tst_1}\sim m_{\tz_1}$, so that top
and stau co-annihilation is possible, while for negative values of
$\alpha$, it appears that co-annihilation is precluded for at least this
choice of parameters. 
\FIGURE[htb]{
\epsfig{file=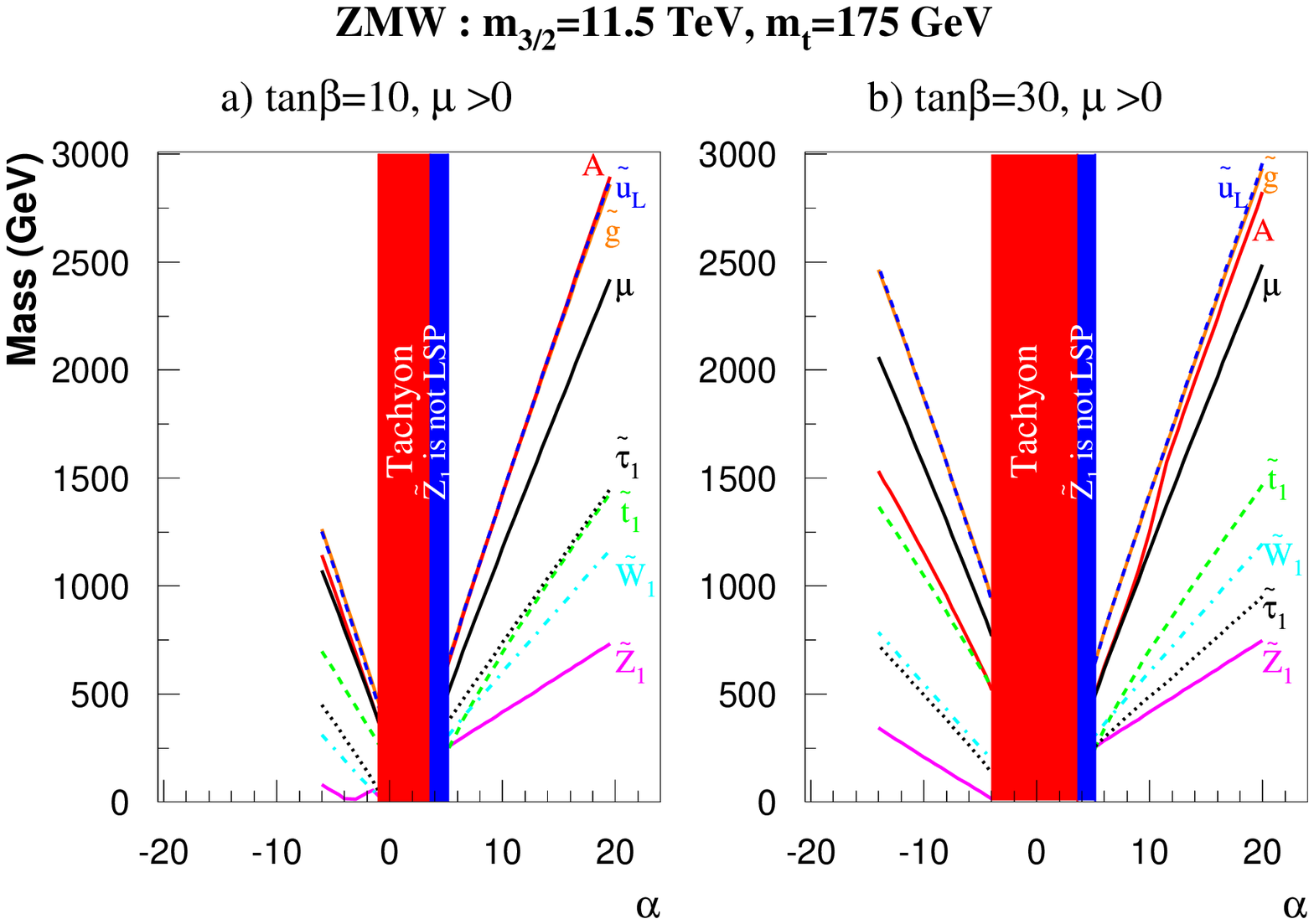,width=12cm} 
\label{fig:mass_zmw_al}
\caption{
Sparticle masses vs. $\alpha$ in the MM-AMSB model with zero modular weights,
for $m_{3/2}=11.5$ TeV and $\mu >0$. We plot for {\it a}) $\tan\beta =10$ and
{\it b}) $\tan\beta =30$.}}

In Fig. \ref{fig:mass_zmw_m32}, we show sparticle masses and the $\mu$
parameter vs. $m_{3/2}$ for fixed $\alpha =6$ and $\mu >0$ and the same
two choices of $\tan\beta$. Along this strip, the neutralino is again
bino-like resulting in too large a relic density.  The exception is
at low $m_{3/2}\sim 8$ TeV where in frame {\it a}) for $\tan\beta =10$
we again find $m_{\tst_1}\sim m_{\tz_1}$, so that top squark
co-annihilation is possible. In frame {\it b}) with $\tan\beta =30$, we
find that for low values of $m_{3/2}$, $m_{\ttau_1}\sim m_{\tz_1}$ so
that tau slepton co-annihilation acts to reduce the relic density (to
too low a value), along
with contributions from the Higgs-funnel annihilation. 

%
\FIGURE[htb]{
\epsfig{file=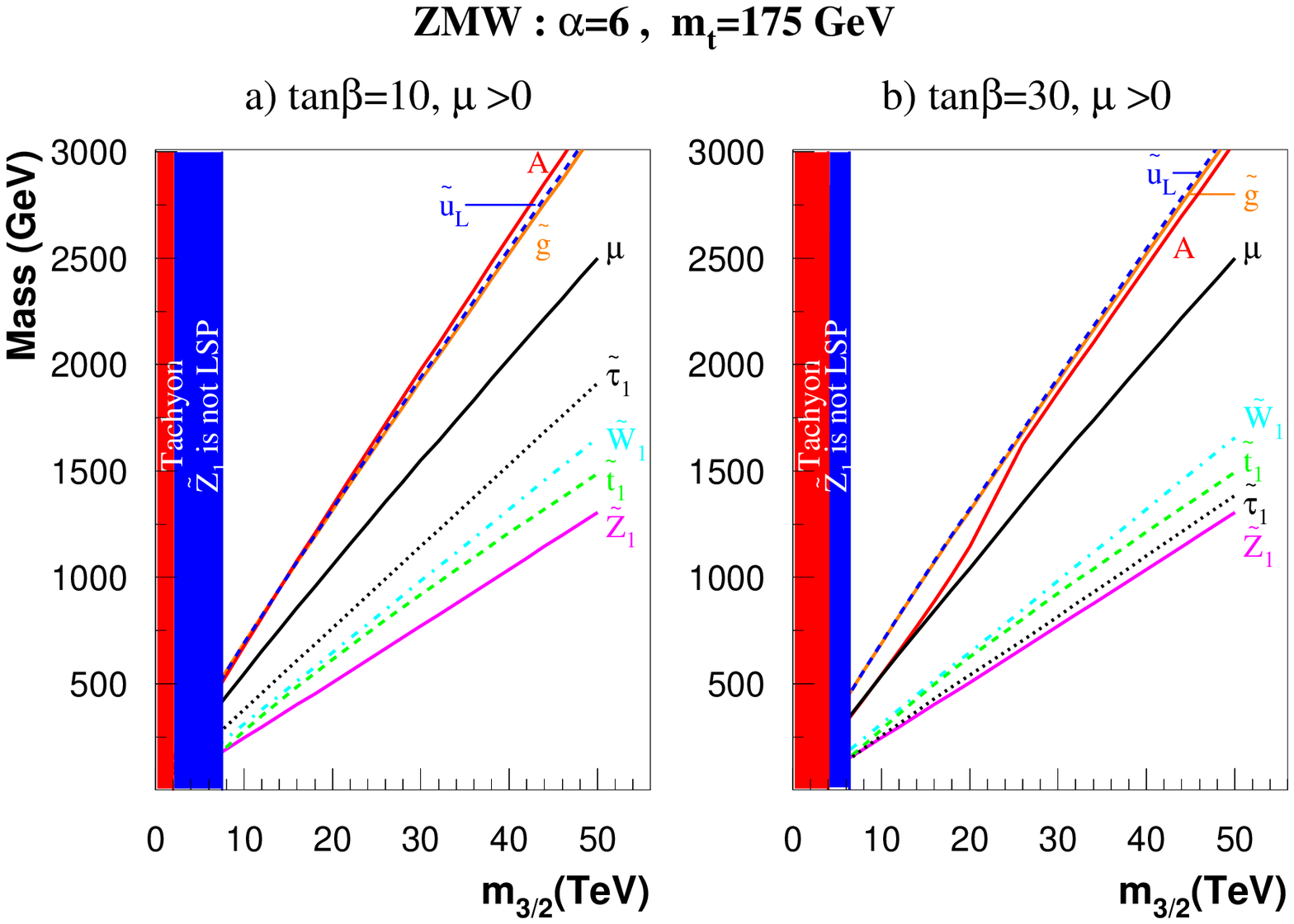,width=12cm} 
\label{fig:mass_zmw_m32}
\caption{
Sparticle masses vs. $m_{3/2}$ in the MM-AMSB model with zero modular weights,
for $\alpha =6$ and $\mu >0$. We plot for {\it a}) $\tan\beta =10$ and
{\it b}) $\tan\beta =30$.}}

In Fig. \ref{fig:mass_zmw_tb}, we plot sparticle masses as well as $\mu$
versus $\tan\beta$ for $\alpha =6$, $m_{3/2}=11.5$ TeV and {\it a}) $\mu
>0$ and {\it b}) $\mu <0$. We see in this case that the allowed
parameter space only extends up to $\tan\beta\sim 31$ for $\mu >0$ and
$\tan\beta \sim 38$ for $\mu <0$. In both cases, the $\ttau_1$ becomes
the LSP at the upper limit on $\tan\beta$, so in this region, $\ttau_1
-\tz_1$ co-annihilation is again possible. Note that the value of $m_A$
is a decreasing function of $\tan\beta$, and for $\mu <0$ (which, for
positive values of $\alpha$, is disfavored by the $(g-2)_\mu$ anomaly), a
region around $\tan\beta\sim 25$ also becomes WMAP allowed, since here
$A$-funnel annihilation can occur, since $2m_{\tz_1}\sim m_A$. The
situation is illustrated in Fig. \ref{fig:Oh2_zmw_tb}, where we plot
$\Omega_{\tz_1}h^2\ vs.\ tan\beta$ for the same parameters as in
Fig. \ref{fig:mass_zmw_tb}.
\FIGURE[htb]{
\epsfig{file=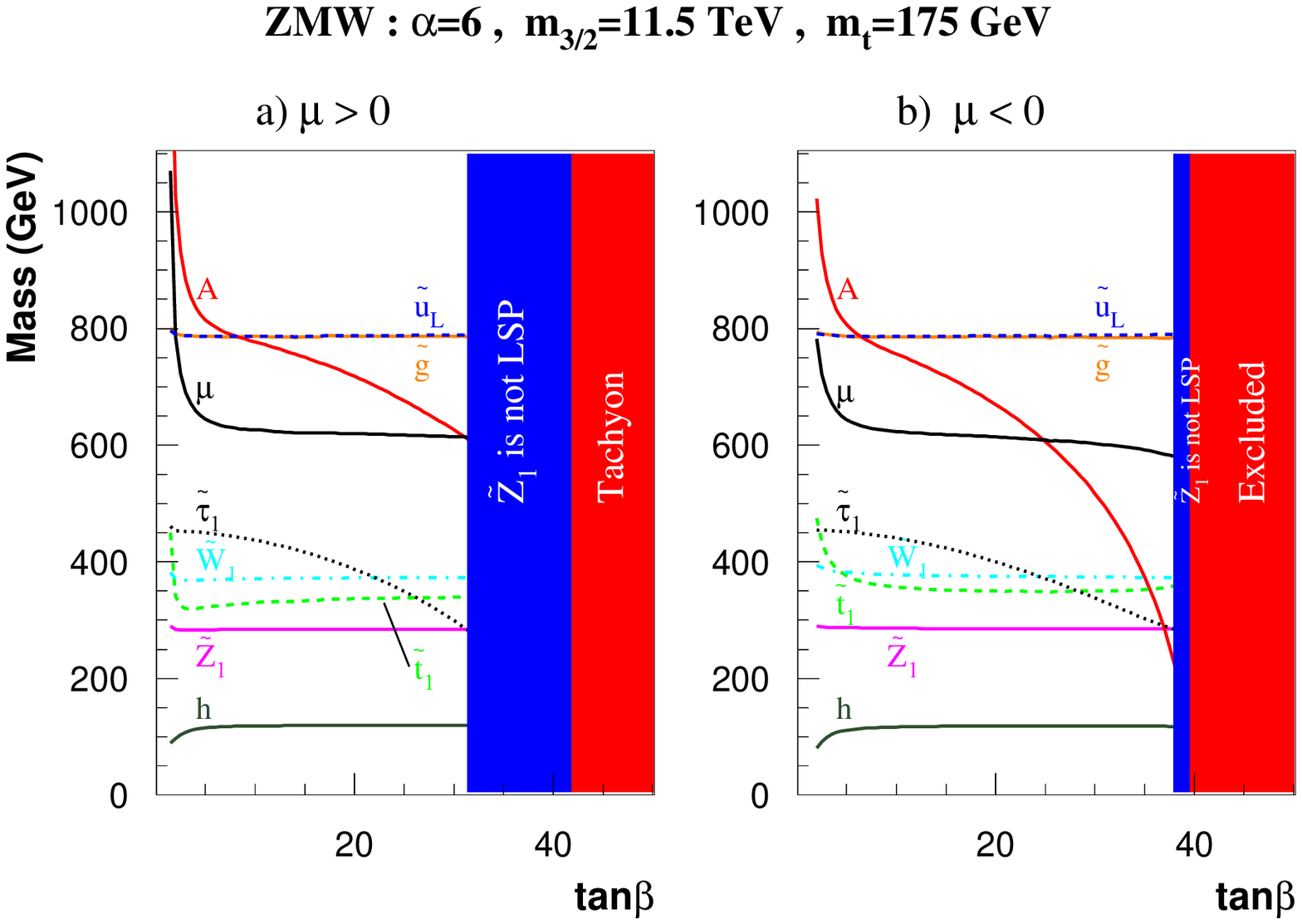,width=12cm} 
\label{fig:mass_zmw_tb}
\caption{ Sparticle masses vs. $\tan\beta$ in the MM-AMSB model with
zero modular weights, for $\alpha =6$ and $m_{3/2}=11.5$ TeV, with
{\it a}) $\mu >0$ and {\it b}) $\mu <0$.}}

\FIGURE[htb]{
\epsfig{file=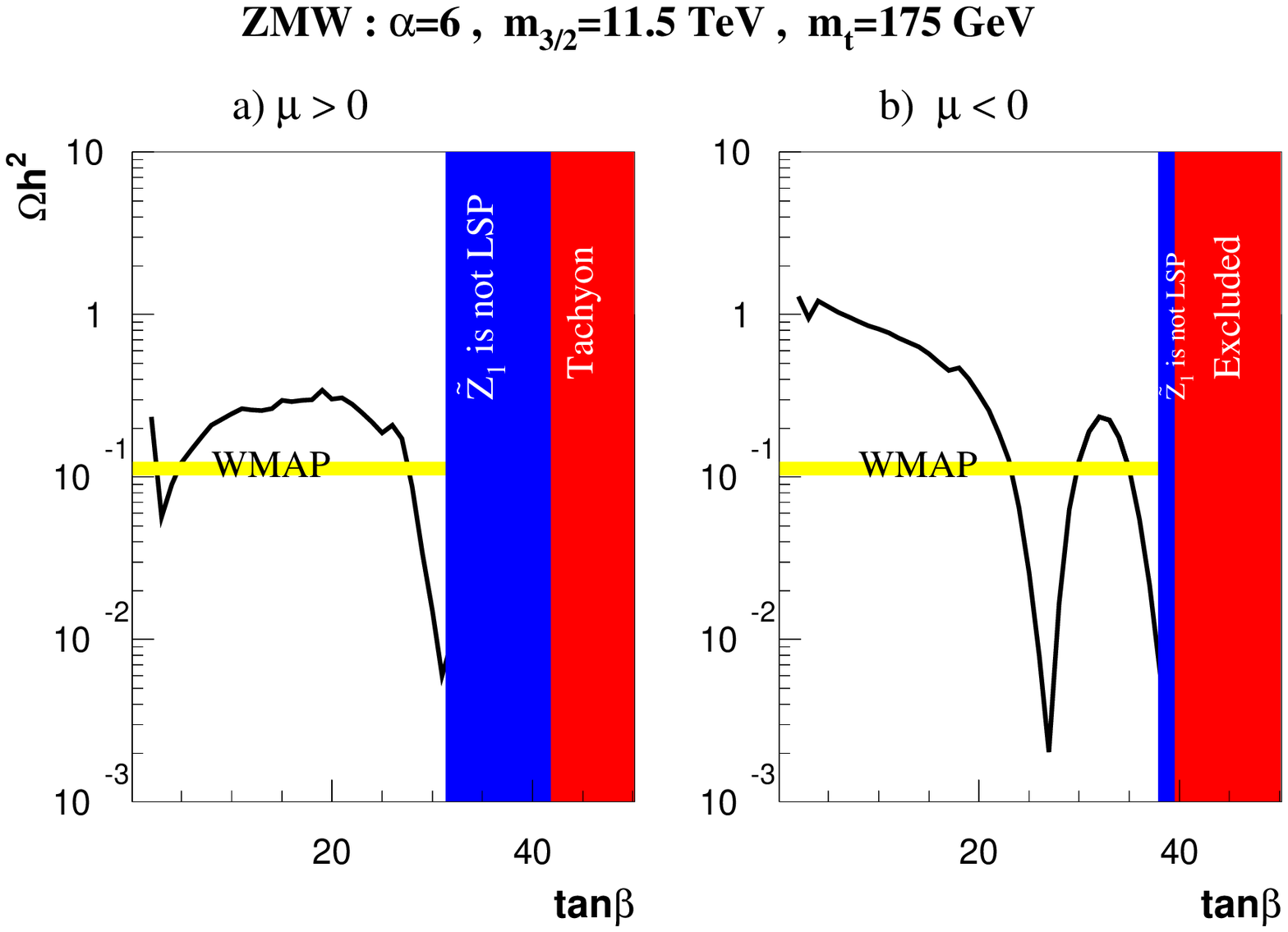,width=12cm} 
\label{fig:Oh2_zmw_tb}
\caption{
The neutralino relic density $\Omega_{\tz_1}h^2\ vs.\ \tan\beta$ for 
the MM-AMSB model with zero modular weights,
for $\alpha =6$ and $m_{3/2}=11.5$ TeV,
with {\it a}) $\mu >0$ and {\it b}) $\mu <0$.}}

In Fig. \ref{fig:zmw_plane}, we show the entire $\alpha\ vs.\ m_{3/2}$
parameter space plane for $\mu >0$ and {\it a}) $\tan\beta =10$ and {\it
b}) $\tan\beta =30$. In the white region, either radiative electroweak
symmetry breaking fails or there are sparticles with masses below the
lower limits from LEP 2. Points with a turquiose square are excluded
because $\tst_1$ is the LSP, while those with a magenta square are
excluded because the LSP is $\ttau_1$. Points denoted by blue dots are
theoretically allowed spectra, but give $\Omega_{\tz_1}h^2>0.5$, and so
are excluded by the relic density measurement.  Points denoted by a
green $\times$ are also excluded, since $0.13<\Omega_{\tz_1} h^2
<0.5$. Finally, points denoted by a red $+$ give $\Omega_{\tz_1}h^2
<0.13$ and are allowed by the relic density constraint. (Remember that
either the CDM could consist of several components, or there may be a
non-thermal component of DM.)  The allowed region extends from very high
values of $m_{3/2}>60$ TeV with $\alpha\sim 5.6$ to very low values of
$m_{3/2}< 3$ TeV for $\alpha >10$; for $\alpha< 0$, the allowed region
extends between 20~TeV $\alt m_{3/2} \alt 35$~TeV.  For positive values
of $\alpha$, we have the bulk of the red $+$s right next to the ``$\tst$
LSP'' region, so that the correct relic density is attained via LSP
co-annihilation with $\tst_1$.  The few red $+$s to the left of the
turquoise $\tst_1$ LSP region at low positive $\alpha$ and large
$m_{3/2}$ are where we have a higgsino-like LSP because $M_1 \sim M_2
\sim M_3$, and the low value of $M_3$ leads to a small value of $\mu$
\cite{belanger,mn,m3dm}. We do not see a region of MWDM where $M_1({\rm
weak}) \sim M_2({\rm weak})$ since the values of $\alpha$ where this
would have occurred lead to $\tst_1$ as the LSP.  Turning to $\alpha <
0$ in frame {\it a}), we see a region consistent with relic density
constraints where $\alpha \sim - 1.5$ and $m_{3/2}\sim 20-35$~TeV.  In
this region, we have $M_1({\rm weak}) \sim -M_2({\rm weak})$ as we
anticipated earlier. For the most negative values of $\alpha\sim -1.63$, 
the relic
density is in accord with (\ref{eq:wmap}) via BWCA \cite{bwca}, while
for somewhat less negative values of $\alpha$, the
LSP becomes wino-like, and an additional source of DM is needed to
saturate the measured value of DM relic density. For the larger value of
$\tan\beta$ in frame {\it b}), the radiative EWSB mechanism fails for
the low negative values of $\alpha$, so that the corresponding region is
absent.\footnote{With large negative values of
$\alpha$, the gaugino masses would be negative and solutions with $\mu >
  0$ would be disfavoured because the SUSY contributions to the muon
  anomalous magnetic moment would be negative. It may be of interest to
  examine this part of the parameter space for solutions with $\mu < 0$.}

\FIGURE[htb]{
\epsfig{file=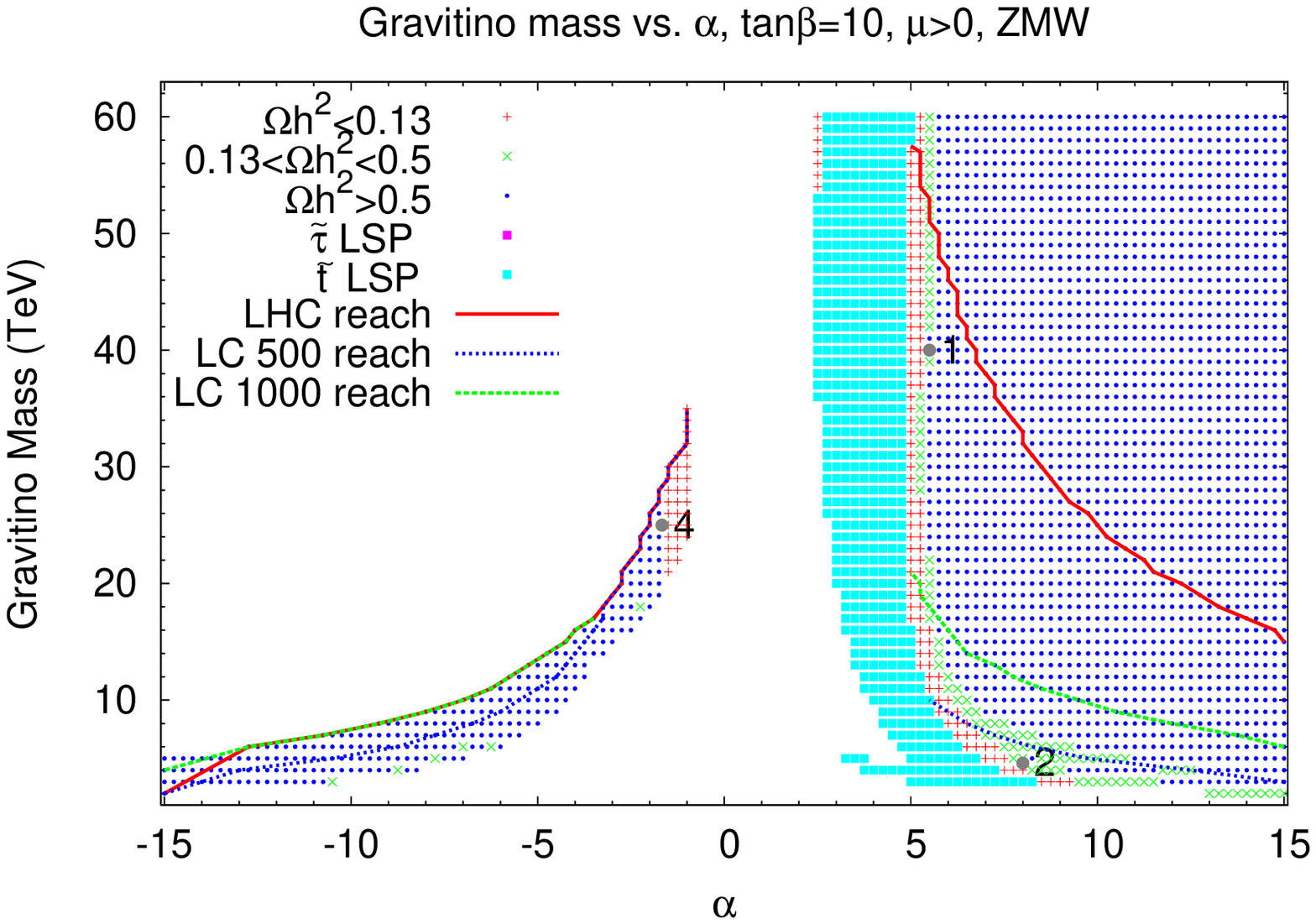,width=12cm} 
\epsfig{file=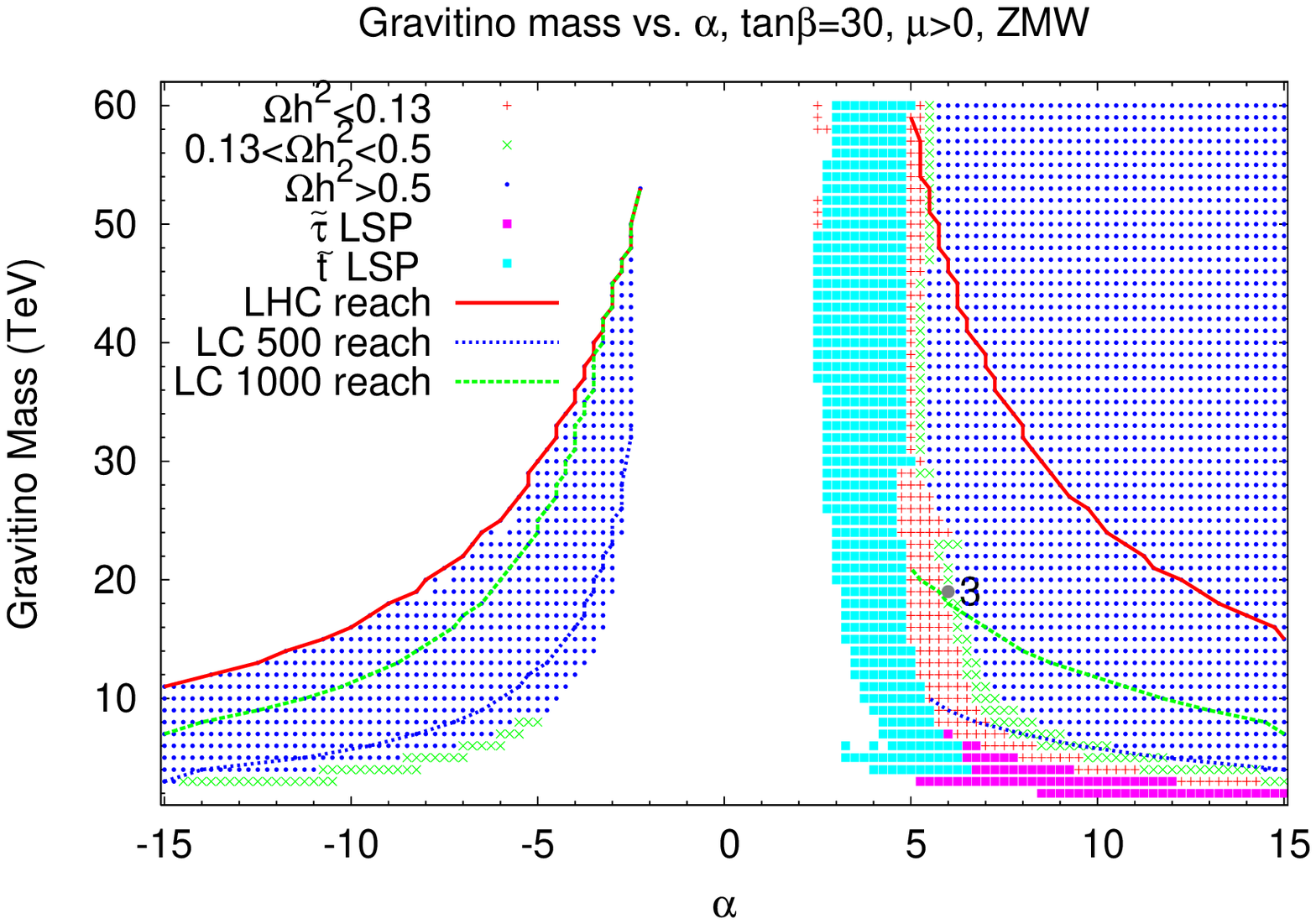,width=12cm} 
\caption{\label{fig:zmw_plane}
Regions of the $\alpha\ vs.\ m_{3/2}$ plane of the MM-AMSB model
where electroweak symmetry is radiatively broken to electromagnetism 
and which are
consistent with experimental constraints from LEP2 searches 
for
{\it a}) $\tan\beta =10$, $\mu >0$ and {\it b}) $\tan\beta =30$, $\mu >0$.
Turquoise (magenta) squares denote points where the $t$-squark (tau
slepton) is the LSP. The ranges of relic density are shown on the
figure. 
We also show the approximate reach of the CERN LHC for
100 fb$^{-1}$ of integrated luminosity, and the kinematic reach of
a $\sqrt{s}=0.5$ and 1 TeV linear $e^+e^-$ collider.}}

Several sparticle masses together with values of various observables
for four sample points are listed in Table \ref{tab:kklt0}. 
Point~1 with large $m_{3/2}=40$ TeV and $\alpha =5.5$ is
characterized by a rather heavy sparticle mass spectrum, but where 
$m_{\tst_1}=1076$ GeV while $m_{\tz_1}=979$ GeV. The 
$\tst_1 - \tz_1$ mass gap is sufficiently low that top squark co-annihilation 
can efficiently reduce the neutralino relic density $\Omega_{\tz_1}h^2$ to
WMAP allowed levels. The  rather heavy sparticle mass spectrum gives rise
to only tiny contributions to $BF(b\to s\gamma )$ and $\Delta a_\mu$
so that these quantities would be expected to be measured at 
nearly their SM values. Point~2 is taken at larger $\alpha =8$ 
but lower $m_{3/2}=4.6$ TeV. This case gives rise to a rather light
sparticle mass spectrum, although the $\tst_1$ is again the NLSP.
A combination of bulk annihilation through $t$-channel sfermion exchange
and top squark co-annihilation act to reduce the relic density to WMAP
allowed levels. The light top squark leads to large non-standard contributions
to $BF(b\to s\gamma )$, and in this case a very low 
branching fraction, below its experimental lower bound would be expected.
%
\begin{table}
\begin{tabular}{lcccc}
\hline
parameter & Point~1 & Point~2 & Point~3 & Point~4 \\
\hline
$\alpha$        & 5.5 & 8  & 6 & -1.635 \\
$m_{3/2} (TeV)$ & 40 & 4.6 & 19 & 25 \\
$\tan\beta$     & 10 & 10 & 30 & 10 \\
$\mu$       & 1753.7 & 371.7 & 967.5 & 961.0 \\
$m_{\tg}$   & 2256.2 & 475.5 & 1257.3 & 1154.4 \\
$m_{\tu_L}$ & 2273.5 & 470.2 & 1261.3 & 1129.4 \\
$m_{\tst_1}$& 1076.5 & 161.4 & 594.4 & 687.2 \\
$m_{\tb_1}$ & 1871.4 & 395.7 & 998.5 & 952.5 \\
$m_{\te_L}$ & 1536.3 & 270.9 & 806.9 & 369.2 \\
$m_{\te_R}$ & 1438.8 & 247.4 & 749.9 & 275.7 \\
$m_{\ttau_1}$ & 1397.8 & 232.0 & 515.0 & 250.4 \\
$m_{\tw_1}$ & 1249.7 & 182.9 & 632.6 & 141.5 \\
$m_{\tz_2}$ & 1245.0 & 183.3 & 630.8 & 141.2 \\ 
$m_{\tz_1}$ & 979.1 & 132.3  & 480.2 & 118.8 \\ 
$m_A$       & 2326.7 & 445.9 & 1076.2 & 987.4 \\
$m_h$       & 124.4 & 114.1 & 122.6 & 116.9 \\
$\Omega_{\tz_1}h^2$& 0.12 & 0.10 & 0.11 & 0.11 \\
$BF(b\to s\gamma)$ & $3.3\times 10^{-4}$ & $9.8\times 10^{-5}$ &
$2.3\times 10^{-4}$ & $4.0\times 10^{-4}$ \\
$\Delta a_\mu    $ & $6.1 \times  10^{-11}$ & $20.1 \times  10^{-10}$ & 
$6.5\times 10^{-10}$ & $-2.2\times 10^{-10}$ \\ 
$BF(B_s \to \mu^+\mu^-)$ &$3.9\times 10^{-9}$ & $4.1\times 10^{-9}$ & 
$5.1\times 10^{-9}$ &$3.8\times 10^{-9}$ \\
$\sigma_{sc} (\tz_1p )$ & $7.9\times 10^{-11}\ {\rm pb}$ & $5.0\times 10^{-9}\ {\rm pb}$ & $2.6\times 10^{-10}\ {\rm pb}$ & $4.6\times 10^{-11}\ {\rm pb}$\\
        $\Phi^{\mu}(km^{-2}yr^{-1})$ & $4\times 10^{-5}$ & $2.61$ & $0.03$ & $10^{-4}$\\
        $\Phi^{\gamma}(cm^{-2}s^{-1})$ & $3.2 \times 10^{-11}\atop(1.6 \times 10^{-15})$ & $2.0\times 10^{-8}\atop(1.0\times 10^{-13})$ & $4.9 \times 10^{-9}\atop(2.5 \times 10^{-13})$ & $2.2\times 10^{-10}\atop(1.0\times 10^{-14})$\\
        $\Phi^{e^+}(GeV^{-1}cm^{-2}s^{-1}sr^{-1})$ & $1.0\times 10^{-12}\atop(2.1\times 10^{-13})$ & $1.3\times 10^{-10}\atop(3.1\times 10^{-11})$ & $1.0\times 10^{-10}\atop(2.3\times 10^{-11})$ & $2.1\times 10^{-11}\atop(4.7\times 10^{-12})$\\
        $\Phi^{\bar p}(GeV^{-1}cm^{-2}s^{-1}sr^{-1})$ & $4.9 \times 10^{-12}\atop(2.8 \times 10^{-13})$ & $1.7\times 10^{-9}\atop(9.6\times 10^{-11})$ & $7.0\times 10^{-10}\atop(3.9\times 10^{-11})$ & $2.9\times 10^{-12}\atop(1.6\times 10^{-13})$\\
        $\Phi^{\bar {D}}(GeV^{-1}cm^{-2}s^{-1}sr^{-1})$ & $1.6\times 10^{-15}\atop(1.4\times 10^{-16})$ & $2.0\times 10^{-12}\atop(1.7\times 10^{-13})$ & $2.9\times 10^{-13}\atop(2.4\times 10^{-14})$ & $5.2\times 10^{-15}\atop(4.4\times 10^{-16})$\\
\hline
\end{tabular}
\caption{Masses and parameters in~GeV units for four case studies of the
MM-AMSB model with zero modular weights. 
 Also shown are predictions for low energy observables, together with
cross sections and fluxes germane to direct and indirect searches
for dark matter. In all cases, we take
$m_t=175$ GeV and $\mu >0$. The halo annihilation rates use the 
Adiabatically Contracted N03 Halo model, while the values in
parenthesis use the Burkert halo profile.}
\label{tab:kklt0}
\end{table}
Point~3 is taken with $\tan\beta =30$, $m_{3/2}=19$ TeV and
$\alpha =6$. While sparticle masses tend to range between 500--1000 GeV, 
we find a stau NLSP with $m_{\ttau_1}=515$ GeV and $m_{\tz_1}=480.2$ GeV.
In this case, it is mainly stau co-annihilation that acts to reduce the 
relic density to WMAP allowed levels. 
This point gives a $BF(b\to s\gamma )=2.3\times 10^{-4}$, somewhat
at the lower end of the range but probably acceptable if we include
theoretical uncertainties for this value of $\tan\beta$.  
Finally, we consider point~4 in the region with negative $\alpha$ where
we get the required relic density via the BWCA mechanism: here, we take
$\alpha= -1.635$, $m_{3/2}=25$ TeV and $\tan\beta=10$. We
see that squark, gluino and Higgs boson masses are qualitatively similar
to those for Point 3, but the uncoloured sparticles are significantly
lighter. The LSP is a bino but the chargino and $\tz_2$ are dominantly
wino-like and close in mass to the LSP. Since the winos can annihilate
efficiently, as long as thermal equilibrium is maintained, the LSP
density is correspondingly reduced. The relatively light $t$-squark and
chargino give a significant contribution to $BF(b \to s\gamma)$, which is
somewhat on the high side. The SUSY contribution to the muon magnetic
moment, though negative, is modest because $\tan\beta$ is only 10.

A rather general feature of the case of the zero modular weight
MM-AMSB model with positive $\alpha$ is the existence of a top squark
with a mass comparable to $m_{\tz_1}$ and a bino-dominated LSP. 
The very light top squark is a consequence of the large $A_t$ parameter
coupled with a value of $M_3$ which is reduced relative to expectations
from models with gaugino mass unification ({\it e.g.} mSUGRA).
Thus, over much of the $\alpha\ vs.\ m_{3/2}$ parameter plane, 
$\tst_1-\tz_1$ co-annihilation leads to a relic density compatible
with the WMAP determination. 
As remarked in Sec.~\ref{sec:intro}, this is in
contrast with the result in Ref.~\cite{flm}, where a low value of
$|\mu|$, and consequently, mixed higgsino dark matter (MHDM) is
obtained because the weak scale $|M_3|$ is reduced relative to $|M_1|$
and $|M_2|$\cite{belanger,mn,m3dm}.
While we do find a reduction in
$|\mu |$, it is not large enough to change the $\tz_1$
from being nearly pure bino to being MHDM.\footnote{The distinction
 between a bino LSP and MHDM is especially important for direct and
 indirect searches for relic DM.} Instead, we find that much
of the phenomenology derives from the rather large magnitude of the $A_t$
parameter in this framework. This, in turn, occurs because
apart from splitting due to the AMSB terms in (\ref{eq:A}), the GUT
scale magnitude of $A_t$
is roughly three times the value of GUT scale gaugino and
scalar masses, and its evolution is approximately flat, resulting in 
a large value also at the weak scale.
Now recall that the RGEs for Higgs and squark SSB mass parameters have
the form, 
\begin{eqnarray}
\frac{dm_{H_d}^2}{dt}&=&{2\over 16\pi^2}\left(-{3\over 5}g_1^2M_1^2-
3g_2^2M_2^2-{3\over 10}g_1^2S+3f_b^2X_b+f_\tau^2X_\tau\right), \\
\frac{dm_{H_u}^2}{dt}&=&{2\over 16\pi^2}\left(-{3\over 5}g_1^2M_1^2-
3g_2^2M_2^2+{3\over 10}g_1^2S+3f_t^2X_t\right), \\
\frac{dm_{Q_3}^2}{dt}&=&{2\over 16\pi^2}\left(-{1\over 15}g_1^2M_1^2-
3g_2^2M_2^2-{16\over 3}g_3^2M_3^2+{1\over 10}g_1^2S+
f_t^2X_t+f_b^2X_b\right), \\
\frac{dm_{\tst_R}^2}{dt}&=&{2\over 16\pi^2}\left(-{16\over 15}g_1^2M_1^2-
{16\over 3}g_3^2M_3^2-{2\over 5}g_1^2S+2f_t^2X_t\right), \\
\frac{dm_{\tb_R}^2}{dt}&=&{2\over 16\pi^2}\left(-{4\over 15}g_1^2M_1^2-
{16\over 3}g_3^2M_3^2+{1\over 5}g_1^2S+2f_b^2X_b\right), 
\end{eqnarray}
where
\begin{eqnarray}
X_t&=&m_{Q_3}^2+m_{\tst_R}^2+m_{H_u}^2+A_t^2 ,\\
X_b&=&m_{Q_3}^2+m_{\tb_R}^2+m_{H_d}^2+A_b^2 ,\\
X_\tau &=&m_{L_3}^2+m_{\ttau_R}^2+m_{H_d}^2+A_\tau^2,\ {\rm and}\\
S&=& m_{H_u}^2-m_{H_d}^2+Tr\left[{\bf m}_Q^2-{\bf m}_L^2-2{\bf m}_U^2
+{\bf m}_D^2+{\bf m}_E^2\right] .
\end{eqnarray}
The large $A_t$, $A_b$ and $A_\tau$ parameters mean that the 
corresponding $X_t$, $X_b$ and $X_\tau$ parameters are also large.
The $f_t^2X_t$ term in $dm_{H_u}^2/dt$ acts to drive $m_{H_u}^2$
to large negative values (this is the well-known REWSB mechanism).
A large negative value of $m_{H_u}^2$ leads to a large $\mu$ value
via the scalar potential minimization condition:
\begin{equation}
\mu^2 =\frac{m_{H_d}^2-m_{H_u}^2\tan^2\beta}{(\tan^2\beta -1)}
-\frac{M_Z^2}{2}\sim -m_{H_u}^2 ,
\end{equation}
where the last equality follows as long as $\tan\beta$ is not very close
to 1, and 
$|m_{H_u}^2 |\gg M_Z^2$.
Thus, even though a lower $M_3$ value acts to reduce $|\mu |$,
the large $A_t$ parameter acts to increase it.
Moreover, the large $X_t$ parameter also acts to {\it suppress}
$m_{\tst_R}^2$ (and $m_{Q_3}^2$) via RG running. 
Finally, the large value of $A_t$ results in
a large intragenerational top squark mixing, and further
reduces the value of $m_{\tst_1}$. 

We thus understand why the MM-AMSB model spectrum, for zero modular
weights and positive $\alpha$, 
is characterized by a bino-like LSP $\tz_1$, but with either a
$\tst_1$ NLSP at lower values of $\tan\beta$, or a $\ttau_1$ NLSP at
high $\tan\beta$. When the mass gap $m_{\tst_1}-m_{\tz_1}$ or
$m_{\ttau_1}-m_{\tz_1}$ is low enough, then co-annihilation can act
reliably to reduce the neutralino relic density to WMAP allowed levels.
Furthermore, in the neighborhood of Point~2, for instance, one can readily
find cases of light $\tst_1$ in this
model with $m_{\tst_1}< m_t$ with $m_h <120$ GeV. These are among some
of the important conditions required for successful electroweak
baryogenesis in the MSSM\cite{ewbg}.

\subsection{Prospects for colliders searches and DM search experiments}

If SUSY is realized as in the MM-AMSB model,  it is expected that
sparticle pair production will occur at observable rates at the CERN
LHC. Indeed, since for positive but not too large values of $\alpha$,
$M_3({\rm weak})$ is {\it smaller} than its value in models with
universal gaugino masses, gluinos (and, via renormalization, also
squarks) will be rather more accessible at hadron colliders within this
framework.  The reach of the LHC for SUSY in the mSUGRA model has been
computed in Ref. \cite{lhcreach}. In these studies, the SUSY production
cross sections are typically dominated by gluino and squark pair
production, followed by cascade decays to a variety of multijet $+\eslt$
plus (multi)-isolated lepton final states. The SUSY reach of the LHC
depends mainly on the gluino and squark masses, and
is relatively insensitive to the particular cascade decay modes which are
active (the details of the cascade decay modes are much more important for
sparticle mass reconstruction and for arriving at the underlying model
parameters). When $m_{\tq}\simeq m_{\tg}$, then the LHC reach with 
100~fb$^{-1}$ of integrated luminosity was found to be $m_{\tg}\sim 3$ TeV,
while in the case where $m_{\tq}\gg m_{\tg}$, the LHC reach extends to
$m_{\tg}\sim 1.8$ TeV.

We expect the LHC reach in the MM-AMSB model to depend mainly on
$m_{\tq}$ and $m_{\tg}$ as well, and not on the particular cascade
patterns. Thus, to obtain the reach within this framework, we simply
digitize the mSUGRA reach contours evaluated in the last of the papers
in Ref. \cite{lhcreach} in terms of $m_{\tg}$ and $m_{\tq}$, and map
them onto the $\alpha\ vs.\ m_{3/2}$ plane of the ZMW model in
Fig. \ref{fig:zmw_plane}. It turns out that for positive $\alpha$,
$m_{\tg}\simeq m_{\tq}$ (while $m_{\tst_1}\sim {1\over 3}-{1\over
2}m_{\tq}$) throughout the parameter half plane, so the LHC reach
extends essentially along the $m_{\tg}\sim 3$ TeV contour line.  
The 100 fb$^{-1}$ LHC reach thus obtained is
shown as the solid red contour in Fig.  \ref{fig:zmw_plane}. The
contour covers all WMAP allowed region when $\alpha$ assumes large
values, but only reaches up to $m_{3/2}\sim 53$ TeV for $\alpha\sim
5.8$. If
$\alpha < 0$, the gluino and squark masses in the allowed region are
well within the reach of the LHC, so that the contour extends
essentially to the
boundary of this allowed region.

We also evaluate the reach of a $\sqrt{s}=0.5$~TeV (1~TeV) linear
$e^+e^-$ collider (LC), where the contour is determined via the
kinematic limits of 0.25 ($0.5$ TeV) for either chargino, stau or stop
pair production.  The corresponding reaches are shown by the blue
(0.5~TeV) and green (1~TeV) contours. 

In the case studies illustrated in Table \ref{tab:kklt0}, we note that
Point~1 is an example of an MM-AMSB model which should yield observable
signals at the LHC via cascade 
decay signatures.  
Gluino and squark pair production, with 
$m_{\tg}\simeq m_{\tq}\simeq 2.3$ TeV, will occur at the LHC with a
cross section of  
several fb. The gluinos mainly decay via two body decays to third
generation quarks and squarks, whereas $\tq_L$ ($\tq_R$) mainly decay
via $\tq_L \to q'\tw_1, \ q\tz_2$ ($\tq_R \to \tz_1$), while $\tw_1\to
\tst_1 b$ and $\tz_2 \to h\tz_1$. Because the mass difference
$m_{\tst_1} - m_{\tz_1}$ is small, the daughter $\tst_1$ will decay via
$\tst_1 \to c\tz_1$: we expect the competing $\tst_1 \to bW\tz_1$ decay
will be more strongly suppressed by phase space. 
Gluino and squark production thus leads to events with 2--4 very hard
jets plus $\eslt$, with an enrichment of $b$-jets, with a not especially
large multiplicity of isolated leptons. 

For Point~2, the very light sparticle mass spectrum will provide
enormous signal rates of order $10^5$ fb, with many multilepton states
and, perhaps, several dilepton mass edges evident. In this case, 
$\tz_2\to e^+e^-\tz_1$ at an enhanced branching fraction level of 5.3\%.
Once again, gluinos mostly decay to third generation quarks and squarks,
but this time, the unusual feature is that the light chargino which is
abundantly produced via cascade decays of $\tq_L$ mostly decays via $\tw_1\to
\tst_1 b$! Indeed, since $m_{\tst_1}$ is only 161~GeV, it may also be
possible to search for $\tst_1$ at the Fermilab Tevatron
\cite{tevatronstop}. 

Point~3 should also lead to readily observable signals at the LHC. Once
again gluinos decay to third generation particles and sparticles,
whereas first generation squark decays are as for Point~1. Since
$\ttau_1$ is relatively light, production of charginos and neutralinos
(either directly, or via cascade decays of $\tq_L$) lead to events rich
in tau jets from $\tz_2\to\tau\ttau$ and 
$\tw_1\to\ttau_1\nu_\tau$ decays, though it should be kept in mind that
$BF(\tw_1 \to \tst_1 b)=47$\%. 

The light $\tst_1$-squark for $\alpha > 0$ results in events rich in
$b$-jets, as we saw for Points~1--3 above. Turning to 
Point~4 with $\alpha <0$, the decay patterns of gluinos and squarks are similar
to those for Points~1--3. The difference is that now, because $\tst_1$
is significantly heavier than $\tz_1$, $\tst_1 \to t\tz_1, \ b\tw_1$ and
even $\tst_1 \to t\tz_2$. The unusual features of this scenario are that
$\tz_2$ dominantly decays to $\tw_1 + f\bar{f'}$ where $f, \ f'$ are
either the quarks of the first two generations or any of the leptons
and neutrinos, and that the chargino mainly decays leptonically via
three body decays with decays $\tw_1 \to \tau\nu_{\tau}\tz_1$ having a
branching fraction of 56\%, while the corresponding decays to each of
the first two generations each occurring 16\% of the time.

Points 1 and 3, while observable at LHC, have no visible two body
sparticle or non-Standard Model Higgs boson processes that will be
accessible at even a 1~TeV LC.  On the other hand, for Point~2 there
should be $\tw_1^+\tw_1^-$, $\tz_1\tz_2$ and $\tst_1\bar{\tst}_1$
signals at a $\sqrt{s}=0.5 $ TeV LC, and possibly even a
$\ttau^+\ttau^-$ signal. Other sleptons, the additional Higgs bosons
and also squarks will be accessible at $\sqrt{s}=1$~TeV.
For Point~4, only $\tw_1$ and $\tz_2$ will be kinematically accessible
at a
$\sqrt{s}=0.5$ TeV machine; however, since the mass gaps between these
and $\tz_1$ is small, specialized analyses \cite{tadaslc} will be
necessary to extract the signal. All the sleptons and sneutrinos should be 
readily accessible at an $e^+e^-$ collider operating at
$\sqrt{s}=1$~TeV. 

Turning to direct and indirect detection of dark matter signals, we
expect these to be in general rather low in this framework, primarily
because the LSP is bino-like, and co-annihilation is usually needed to
reduce the relic density. We have computed direct (using
Isatools\cite{isatools}) and indirect detection rates (using
DarkSUSY\cite{darksusy}) for the four case studies in Table
\ref{tab:kklt0}, and list these in the last few rows. Second generation
direct detection experiments such as CDMS2 expect to probe
spin-independent neutralino-proton scattering cross sections down to
$\sim 3\times 10^{-8}$ pb, while Stage 3 experiments such as SuperCDMS,
Xenon or Zeplin4 aim to reach the $10^{-9}$ pb level. We see from
Table~\ref{tab:kklt0} that none of the Points~1--4 will yield an
observable signal at Stage 2 detectors, while just Point~2 will be
detectable at a Stage 3 detector.  Likewise, neutrino telescopes such as
IceCube are expected to probe neutralino annihilation to SM particles in
the core of the sun.  The SM particles will produce high energy
neutrinos, which can be detected in polar ice via conversion to muons at
the level of 40 events/km$^2$/yr, for $E_\mu >50$ GeV.  With this
criterion, none of the points in the Table will be visible at IceCube.

Neutralino dark matter can also be detected via annihilations in the
galactic core or halo to $\gamma$s, $e^+$s, $\bar{p}$s or anti-deuterons
$\bar{D}$\cite{indirect}.  We focus on gamma ray signals emanating from
the galactic core since the signal is expected to be the largest in this
direction.  In this case, the gamma rays arise from neutralino
annihilation to SM particles in the galactic core, where the SM
particles hadronize to pions, and $\pi^0\to \gamma\gamma$. In this case,
we expect a continuous signal distribution where 1 GeV$<E_\gamma
<m_{\tz_1}$ (the lower limit comes from the minimum energy we require
for the GLAST detector). In fact, the upper cut-off in the gamma energy
distribution, if observed, would give a measure of the WIMP mass. It
should, however, be kept in mind that HESS\cite{hess} and
MAGIC\cite{magic} experiments have detected TeV gamma rays with energies
ranging from their detectability threshold $\sim 200$~GeV to around
10~TeV from the galactic center. The lower end of this range
overlaps with the energies of the gamma rays expected from neutralino
annihilation because, in the scenarios we considered here, $m_{\tz_1}$
typically ranges between 100-1000 GeV. Thus any LSP signal from the
galactic center will have to be identified as an excess above this continuum, 
with a cut off in its energy spectrum at $m_{\tz_1}$. For
LSP annihilation from regions away from the center of our Galaxy, the
gamma rays detected by HESS and MAGIC experiments will not be an issue, but
the signal may be smaller.
 
The GLAST experiment expects to be able to detect $\gamma$s at the
$10^{-10}$ events/cm$^2$/sec level for $E_\gamma >1$ GeV. Using this
criterion, and the Adiabatically Contracted N03 Halo model\cite{n03}, we
find that Points~2--4 might find detectable signals from the direction
of the galactic center. It should be remembered though that the gamma
ray flux is sensitive to the assumed halo profile, and that our
assumption of the Adiabatically Contracted N03 Halo model gives us an
optimistic projection for the rate.  The rates for indirect dark matter
detection using an alternative less optimistic halo density profile (the
Burkert profile\cite{burkert}) are listed in parenthesis in the table
below the rates using the N03 profile.  Using the Burkert profile, none
of the points would be detectable via gamma rays from the galactic
center.

Turning to antimatter experiments, we compute the solar-modulated 
positron, antiproton and antideuteron fluxes, following the procedure 
outlined in Ref.~\cite{Profumo:2004ty}. 
We calculate the neutralino annihilation rates to $\bar{p}$ and
$\bar{n}$ 
using the 
{\tt Pythia} 6.154 Monte Carlo code~\cite{pythia} as implemented in 
{\tt DarkSUSY}~\cite{darksusy}, and then deduce the $\overline{D}$ 
yield using the prescription suggested in Ref.~\cite{dbar}.
For the propagation of charged 
cosmic rays through the galactic magnetic fields, we use the default DarkSUSY
model where propagation is worked out through 
an effective two-dimensional diffusion model in the steady state 
approximation. We refer the reader to the DarkSUSY manual for more 
details\cite{darksusy}. Solar modulation effects are implemented through the 
analytical force-field approximation of 
Gleeson and Axford~\cite{GleesonAxford}.

For positrons and anti-protons we evaluate the averaged differential 
antiparticle flux in a projected energy bin centered at a kinetic 
energy of 20 GeV, where we expect an optimal statistics and 
signal-to-background ratio at space-borne antiparticle 
detectors\cite{antimatter,statistical}. 
We take the  
experimental sensitivity to be that of the Pamela experiment 
after three years of data-taking as our benchmark:
$2\times 10^{-9}$ events/GeV/cm$^2$/sec/sr for positrons, and
$3\times 10^{-9}$ events/GeV/cm$^2$/sec/sr for anti-protons.
We find that, even with the optimistic assumption of the N03
halo profile,  none of the points will yield an observable signal in
the  Pamela experiment via positrons or anti-protons. 

Finally, the average differential antideuteron flux has been computed in
the $0.1<T_{\bar D}<0.25$ GeV range, where $T_{\bar D}$ stands for the
antideuteron kinetic energy per nucleon, and compared to the estimated
GAPS sensitivity for an ultra-long duration balloon-borne experiment
\cite{gaps} (see Ref.~\cite{baerprofumo} for an updated discussion of
the role of antideuteron searches in DM indirect detection).  With a
projected GAPS sensitivity of $3\times 10^{-13}$
events/GeV/cm$^2$/sec/sr, just Point~2 may lead to an 
observable signal with the N03 halo profile,
though the signal from Point~3 is right on the edge of detectability.

While the direct or indirect detection of DM may be difficult or
impossible for most of the parameter range in this scenario, primarily
because the LSP is bino-like, we should
keep in mind that there are regions where agreement with (\ref{eq:wmap})
may be obtained with parameters in the $A$-funnel\cite{indirect}; see
Fig.~\ref{fig:Oh2_zmw_tb}. In this case, we may expect that detection becomes
possible via direct detection experiments, and via indirect
anti-particle and gamma ray searches, but not in IceCube (because
the spin-dependent LSP nucleon cross section is not enhanced).

\section{MM-AMSB Model with non-zero modular weights}
\label{sec:nzmw}

The results of the previous section were all obtained in MM-AMSB model
where the modular weights $n_i$ were all taken to be zero.  Many
other choices of modular weights may be taken depending on which branes
the matter and Higgs fields inhabit. Each choice will yield a somewhat
different low energy phenomenology, since the soft terms all depend on
the modular weights.  In this section, we will illustrate how the
phenomenology changes for a particular choice: we will retain $\ell_a=1$,
but take Higgs fields to live in a $D3$ brane, so that
$n_{H_u}=n_{H_d}=1$, while matter fields live on intersections of $D7$
branes, so that $n_{\rm matter}={1\over 2}$.  We will, for brevity,
refer to this as the non-zero modular weights (NZMW) model.  We have
also examined the choice $n_{\rm matter}=1$, $n_{H_u}=n_{H_d}=0$. We
will discuss this case briefly at the end of Section~\ref{subsec:nzmwspec}.

\subsection{Soft SUSY breaking terms}

While the gaugino SSB mass parameters are unaltered, the soft SUSY
breaking terms for the scalars are modified from the values
presented in Fig. \ref{fig:soft_zmw}.  In general, a non-zero choice for
the modular weights reduces importance of the modulus-mediated
contributions to the SSB masses and $A$-parameters relative to the AMSB
contributions.  For our particular choice $n_{H_u}=n_{H_d}=1$ and
$n_{\rm matter}={1\over 2}$, the $a_{ijk}$ coefficients of the $\alpha$ term
in Eq. \ref{eq:A} are diminished, so that the common GUT scale value
of the $A-$parameters is now $\sim -M_s$ rather than $-3M_s$, and these no
longer dominate $X_t$ as in the previous section.
In addition, the $c_i$
coefficients in Eq. \ref{eq:m2} are diminished, which enhanced the AMSB
and mixed modulus-anomaly mediated contributions to the soft scalar
squared masses.  

As an example, we plot in Fig. \ref{fig:soft_nzmw} the same soft
parameters as in Fig. \ref{fig:soft_zmw}, but for the NZMW case.  In
frame {\it a}), the gaugino masses are of course unaffected. However,
the GUT scale values of trilinear $A$-parameters are substantially
reduced in magnitude compared to Fig. \ref{fig:soft_zmw}{\it a}). Thus,
we expect in the case of the NZMW model a reduced diminution of the top
squark soft terms, and hence a heavier $\tst_1$.  In frame {\it b}), we
see the third generation and Higgs scalar masses.  Taking
$n_{H_u}=n_{H_d}=1$ means $c_{H_u}=c_{H_d}=0$, so that the pure
modulus-mediated contribution to Higgs squared masses is absent. In
fact, the Higgs squared masses have negative values for the entire range
of $\alpha$ shown.  In addition, the modulus-mediated contribution to
the squark and slepton soft masses is diminished compared to the zero
modular weights case.  This results in the mixed anomaly/modulus
mediation contribution dominating the soft masses for a larger range of
$\alpha$ than in the case of zero modular weights, so that GUT scale
soft squared masses are also negative over a range of $\alpha$ values.
As before, after renormalization group
evolution to low scales, we see that over a portion of this range of
$\alpha$ we find acceptable spectra at the weak scale.

%
\FIGURE[htb]{
\epsfig{file=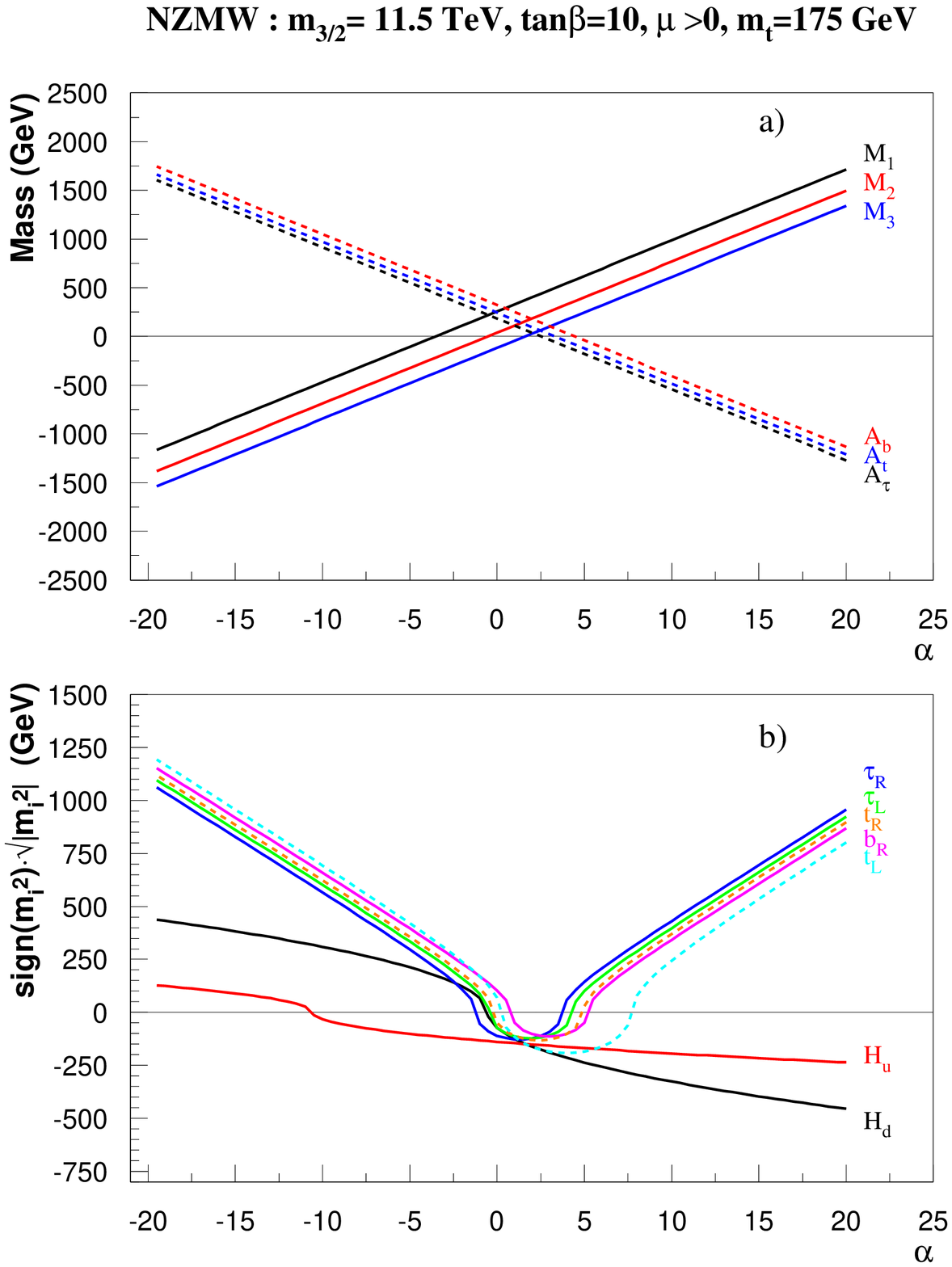,width=12cm} 
\caption{\label{fig:soft_nzmw}
Various soft SUSY breaking parameters at the scale
$Q=M_{GUT}$ versus $\alpha$ for the NZMW model with $n_{\rm matter}= {1\over 2}$, 
$n_{Higgs}=1$, $\ell_a =1$, 
$m_{3/2}=11.5$ TeV, $\tan\beta =10$, $\mu >0$ and $m_t=175$ GeV.
In {\it a}), we show gaugino masses and $A$ terms, while in 
{\it b}) we show $sign(m_i^2)\cdot\sqrt{|m_i^2|}$ for third generation and
Higgs soft breaking masses.}}

In Fig. \ref{fig:mi_evol_nz}{\it a}) we show the evolution of gaugino
masses for the case $\alpha =6$, $m_{3/2}= 11.5$ TeV, $\tan\beta =10$
and $\mu >0$. They evolve in essentially the same fashion as
Fig. \ref{fig:mi_evol} (since the effects of modular weights only enter
either via decoupling in the RGEs, or via two loop terms in the RGE),
and again show mirage unification at $Q\sim 10^{11}$ GeV. The $A_t$,
$A_b$ and $A_\tau$ evolution is shown in frame {\it b}). In this case,
these parameters evolve to weak scale values which are comparable to the
other soft masses.  While they do not show exact mirage unification
owing to Yukawa coupling effects in their evolution, these apparently
unify much better than in the case of the ZMW model because the Yukawa
coupling terms in the evolution of the $A$-parameters are now smaller because
of the reduced values of the $A$-parameters.
\FIGURE[htb]{
\epsfig{file=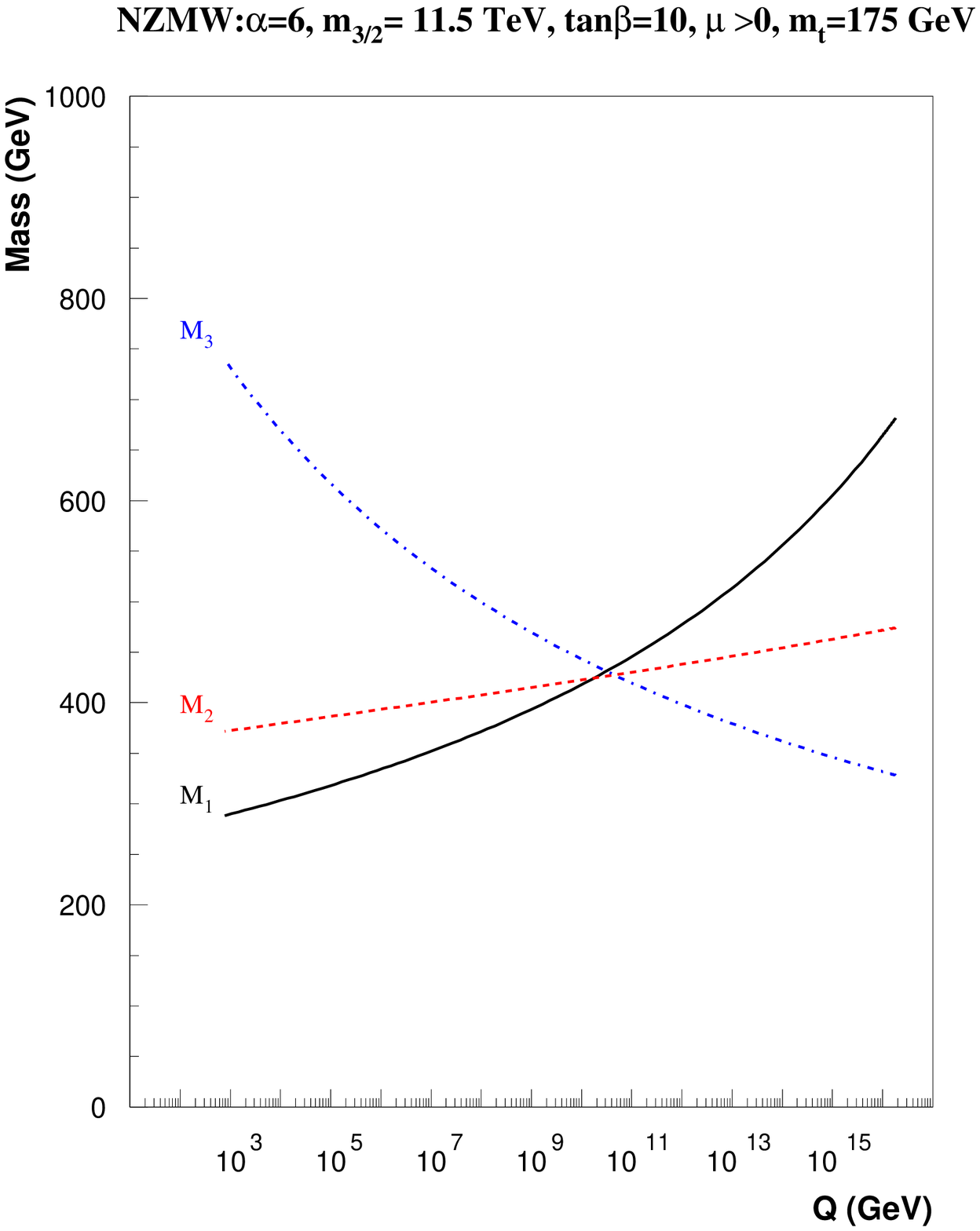,width=7cm} 
\epsfig{file=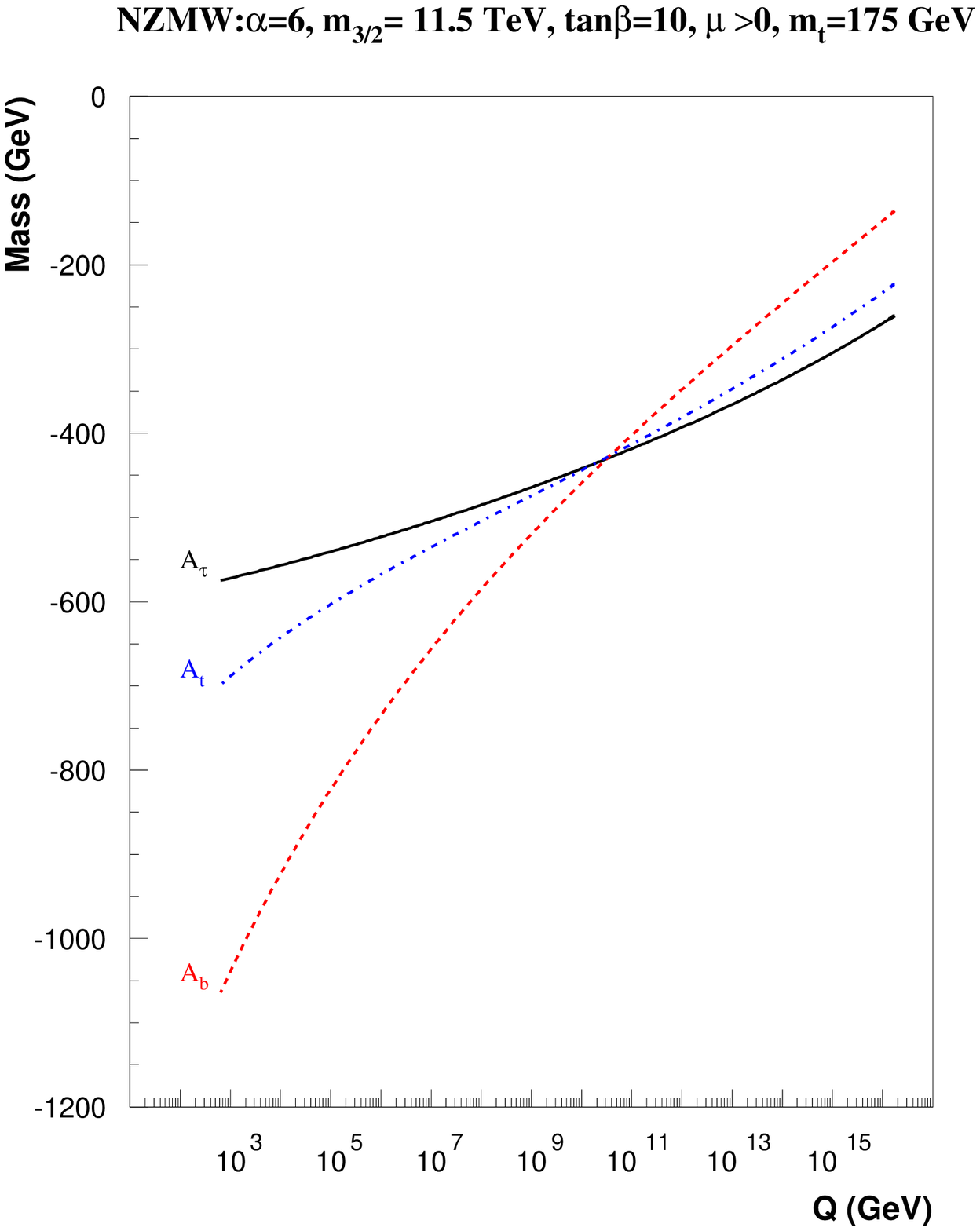,width=7cm} 
\caption{\label{fig:mi_evol_nz}
Evolution of {\it a}) the gaugino masses $M_1$, $M_2$ and $M_3$ from
$Q=M_{GUT}$ to $Q=M_{\rm weak}$ in the NZMW model for $\alpha =6$,
$m_{3/2}= 11.5$ TeV, $\tan\beta =10$, $\mu >0$ and $m_t=175$ GeV.
In frame {\it b}, we show evolution of the $A_i$ trilinear 
soft masses $A_t$, $A_b$ and $A_\tau$ from $Q=M_{GUT}$ to $Q=M_{\rm weak}$ for
the same parameter choices.
}}

In Fig. \ref{fig:sm_evol_nz}, we show the evolution of scalar SSB mass
parameters from $Q=M_{GUT}$ to $Q=M_{\rm weak}$, for the same model choice
as in Fig. \ref{fig:mi_evol_nz}. In this case, it is interesting to note
that $m_{H_u}^2$ first evolves from negative to positive values, and
then back to negative again, {\it i.e} it would have been premature to
conclude that electroweak symmetry is broken at the tree-level from the
fact that the Higgs boson squared mass parameters were negative. We also
see that though several matter SSB mass squared parameters are negative
at the GUT scale, they ultimately evolve to positive values, and we
obtain acceptable weak scale spectra.  This re-iterates comments made in
Sec.~\ref{sec:zmw} about the importance of the radiative corrections to
the potential with parameters renormalized at a scale much higher than
the weak scale.  Note that the matter scalars and the Higgs scalars now
show separate (approximate) mirage unification at
the common scale of $Q\sim 10^{11}$~GeV.
\FIGURE[htb]{
\epsfig{file=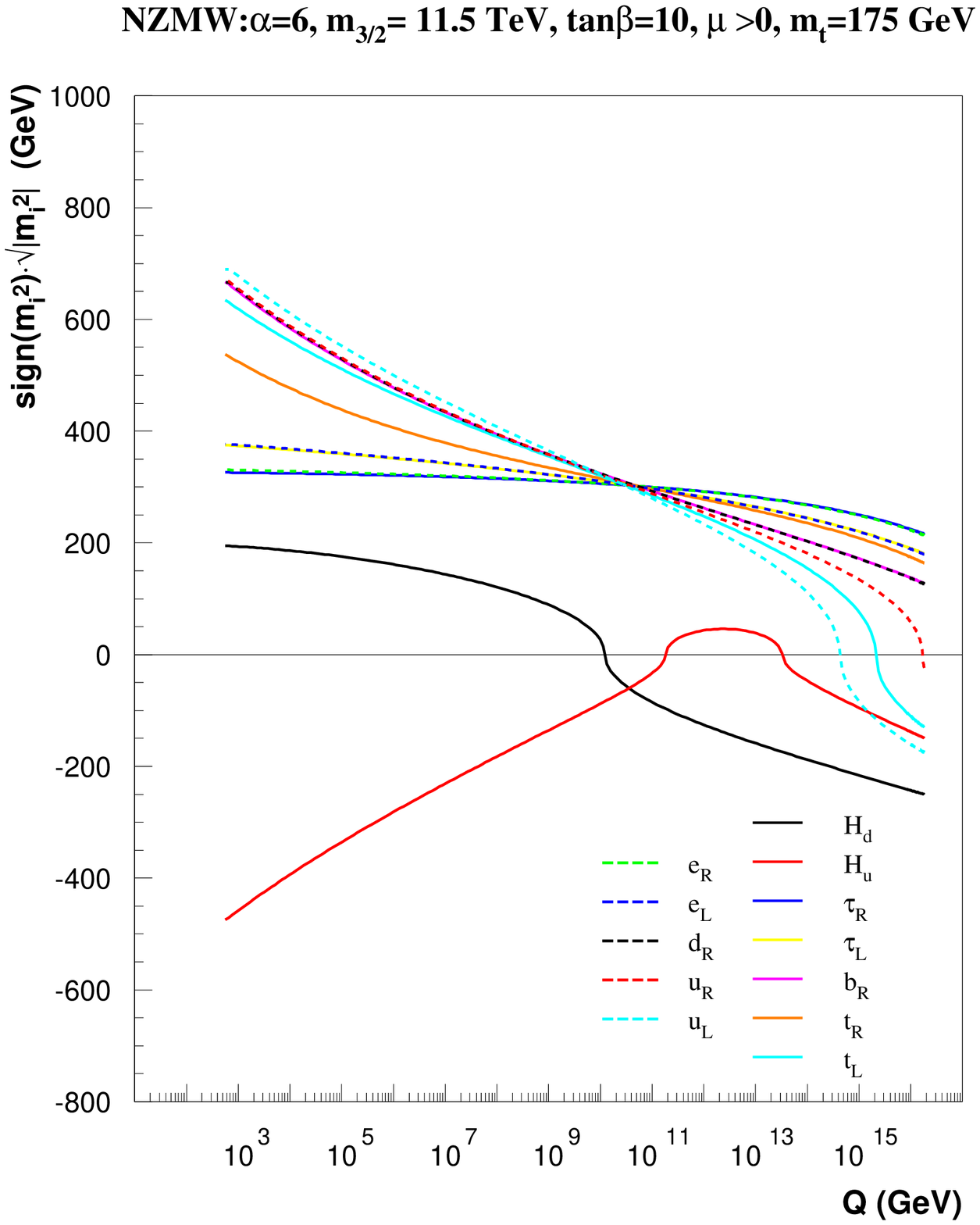,width=10cm} 
\caption{\label{fig:sm_evol_nz}
Evolution of scalar soft masses $m_i^2$ for {\it a}) first generation
scalars and {\it b}) third generation and Higgs scalars from
$Q=M_{GUT}$ to $Q=M_{\rm weak}$ for $\alpha =6$,
$m_{3/2}=11.5$ TeV, $\tan\beta =10$ and $\mu >0$ for the case
of non-zero modular weights.
We actually plot $sign(m_i^2)\cdot\sqrt{|m_i^2|}$.}}

\subsection{Mass spectrum} \label{subsec:nzmwspec}

In Fig. \ref{fig:mass_nzmw_al}, we plot the physical sparticle masses
versus $\alpha$ for $m_{3/2}=11.5$ TeV, $\mu >0$ and {\it a}) $\tan\beta
=10$ and {\it b}) $\tan\beta =30$.  In both cases, $|\mu |$ is large
over most of the range of $\alpha$, resulting in a bino-like LSP.  Also,
in both cases, the $\ttau_1$ is the NLSP, and in fact the lower range of
(positive) $\alpha$ is bounded by the requirement of an uncharged LSP.
A striking feature is the narrow allowed sliver of $\alpha > 0$ between
the red and blue regions in frame {\it b}). In this region, $\mu$ also
becomes very small, and mixing between the bino and the higgsino
depresses the LSP mass below $m_{\ttau_1}$, and the LSP is a roughly
equal mixture of bino, wino and the higgsinos.
We also see that the $\tst_1$ is seen to be typically much more massive
than the $\tz_1$, so that in this case top squark co-annihilation plays
no role in reducing the relic density. However, there does exist a range
of $\alpha$ values for which $2m_{\tz_1}\sim m_A$ ($\alpha \sim 6-8$ in
frame {\it a}) and $\alpha \sim 8-11$ in frame {\it b})), so that
$A$-funnel annihilation can act to reduce the relic density over some
range of parameter choices.
\FIGURE[htb]{
\epsfig{file=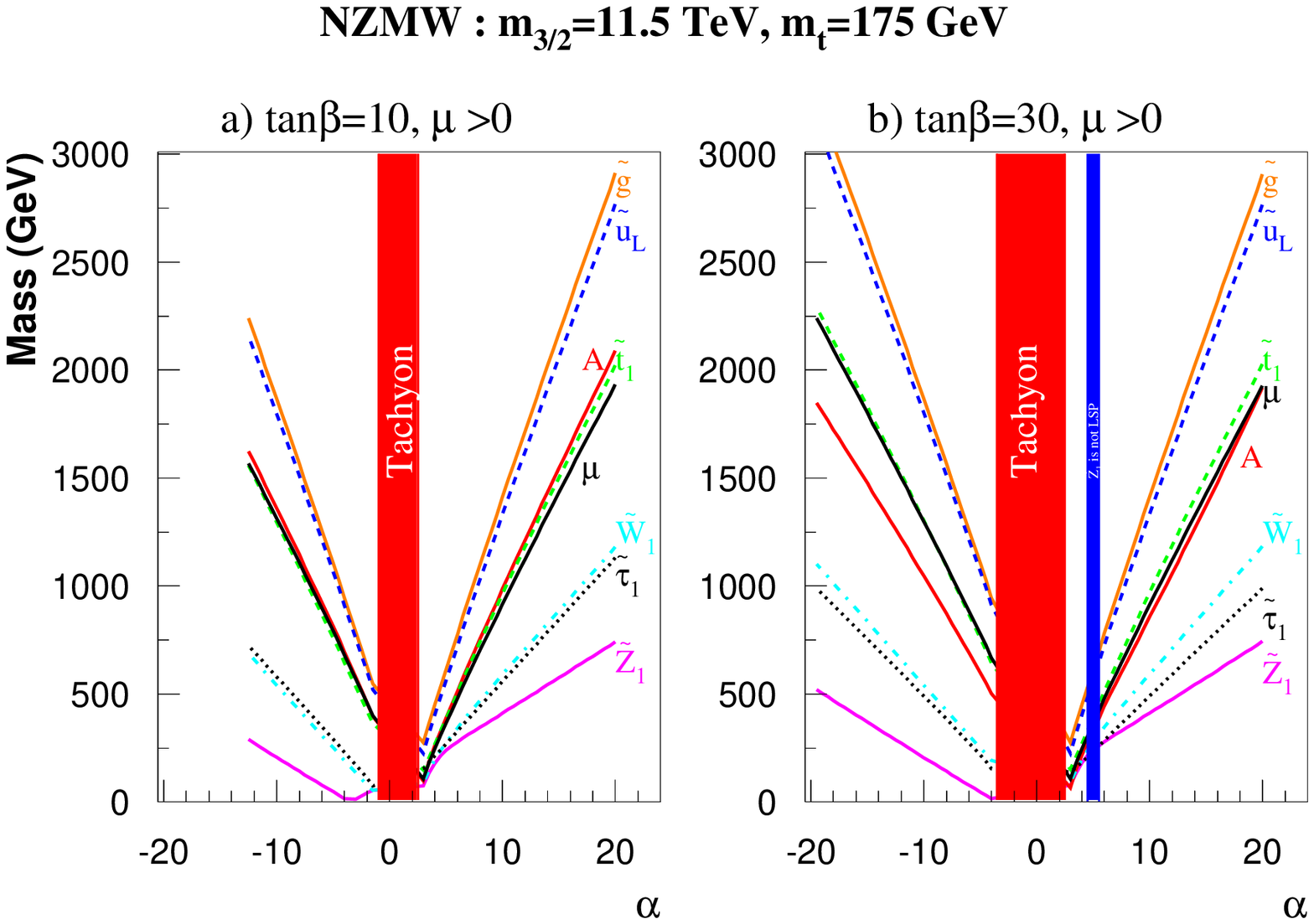,width=12cm} 
\label{fig:mass_nzmw_al}
\caption{
Sparticle masses vs. $\alpha$ in the NZMW  model,
for $m_{3/2}=11.5$ TeV and $\mu >0$. We plot for {\it a}) $\tan\beta =10$ and
{\it b}) $\tan\beta =30$.}}

In Fig. \ref{fig:mass_nzmw_tb}, we plot the physical sparticle masses
versus $\tan\beta$ for $\alpha =6$, $m_{3/2}=11.5$ TeV, and {\it a})
$\mu > 0$ and {\it b}) $\mu < 0$.  In this case, again the magnitude of
$\mu$ stays large, while the upper bound on $\tan\beta$ comes from the
requirement of a neutralino LSP.  In both frames, we see that there is a
range a $\tan\beta$ where
$2m_{\tz_1}\sim m_A$, so that there is efficient resonant annihilation
of relic neutralinos. Notice that $m_{\tw_1}:m_{\tz_1}$ differs
significantly from $2:1$ expected in models with gaugino mass
unification. 
\FIGURE[htb]{
\epsfig{file=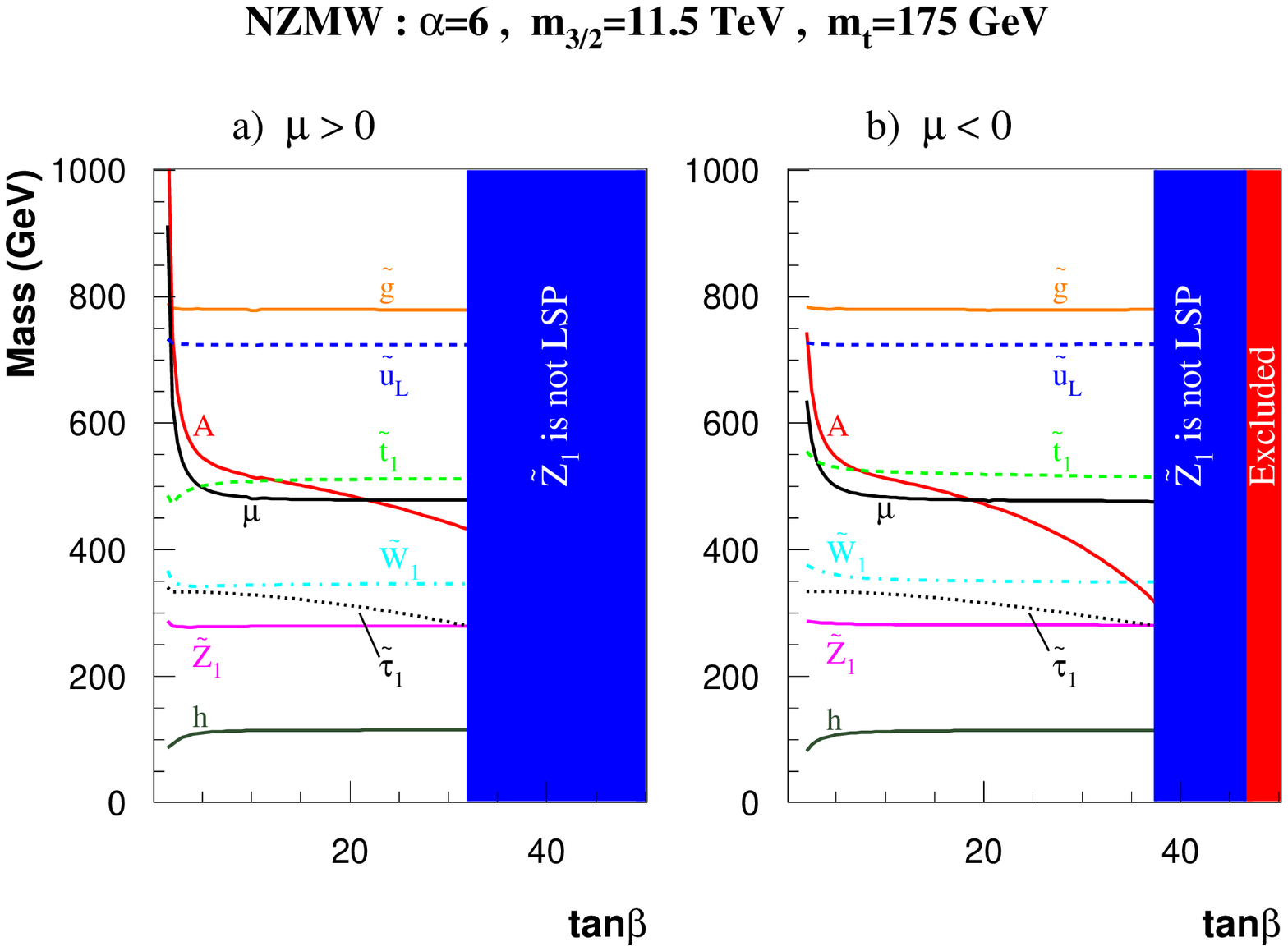,width=12cm} 
\label{fig:mass_nzmw_tb}
\caption{
Sparticle masses vs. $\tan\beta$ in the NZMW model
for $\alpha =6$ and $m_{3/2}=11.5$. 
We plot for {\it a}) $\mu >0$ and {\it b}) $\mu < 0$.}}

The relic density for the NZMW model for the same parameters as in
Fig.~\ref{fig:mass_nzmw_tb} is shown in Fig.~\ref{fig:Oh2_nzmw_tb}. In
both frames  stau co-annihilation reduce
the relic density at
the upper end of the range of $\tan\beta$, while at the lower value of
$\tan\beta$, this
reduction occurs via $s$-channel resonant annihilation via $A, \ H$
bosons. 

\FIGURE[htb]{
\epsfig{file=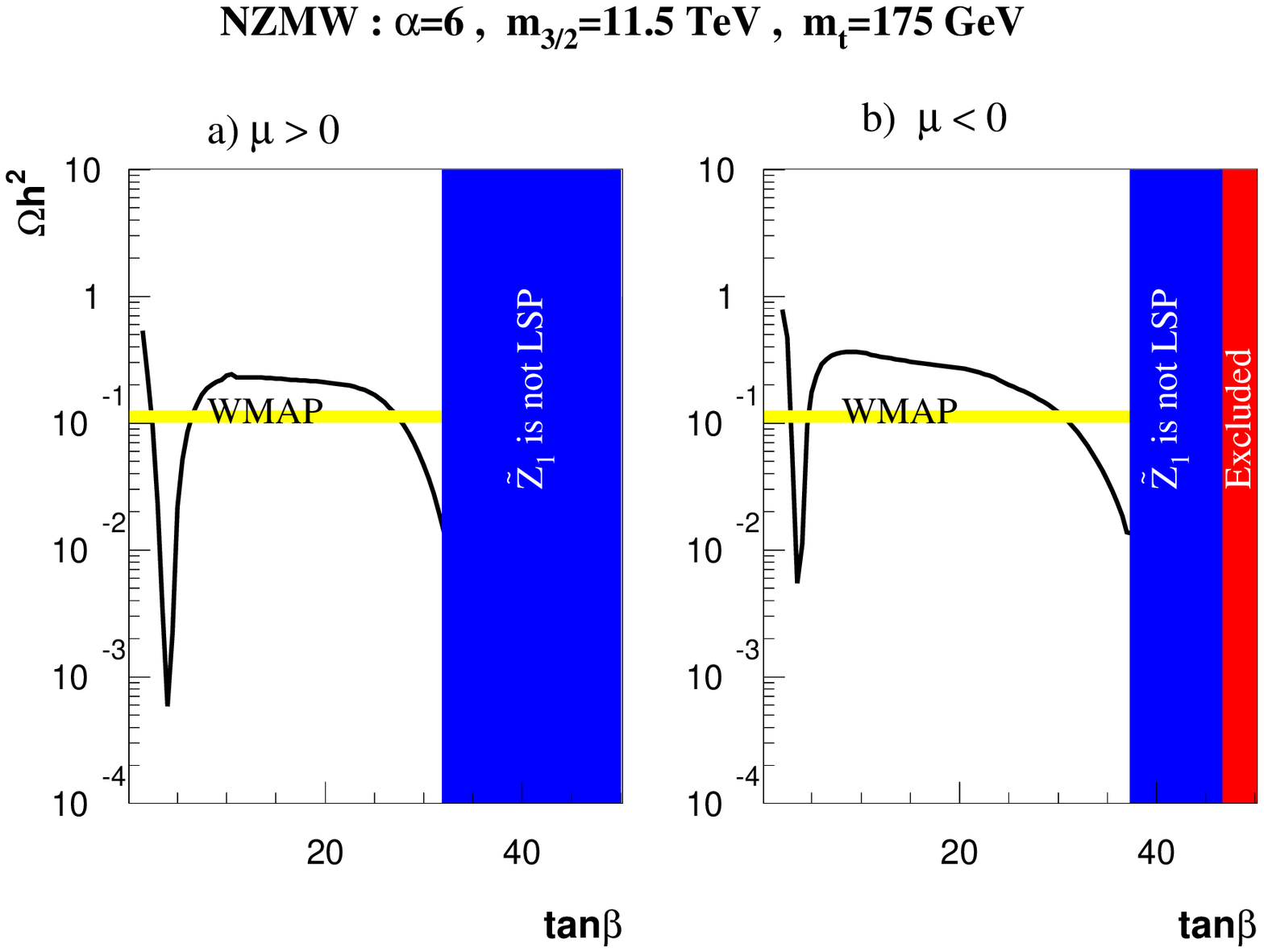,width=12cm} 
\label{fig:Oh2_nzmw_tb}
\caption{
The relic density $\Omega_{\tz_1}h^2\ vs.\ \tan\beta$ for 
the NZMW model with
$\alpha =6$ and $m_{3/2}=11.5$. 
We plot for {\it a}) $\mu >0$ and {\it b}) $\mu <0$.}}

In Fig.~\ref{fig:nzmw_plane}, we plot the allowed regions of the MM-AMSB
model with non-zero modular weights in the $\alpha\ vs.\ m_{3/2}$ plane
for $\mu > 0$ and {\it a}) $\tan\beta =10$ and {\it b}) $\tan\beta =30$.
The notation used is the same as in Fig. \ref{fig:zmw_plane}.  Aside
from the white region where either the theoretical constraints or the
bounds from LEP 2 searches are not satisfied, portions of the plane are
excluded when the $\ttau_1$ becomes the LSP. This is in contrast to the
ZMW model where large parts of this plane were excluded because $\tst_1$
becomes the LSP.  The boundary between the green $\times$ and red $+$
regions gives $\Omega_{\tz_1}h^2\sim 0.11$.  For positive values of
$\alpha$, both frames are qualitatively similar: for large $m_{3/2}$
with $\alpha\sim 5$ or 6, the WMAP allowed region occurs due to stau
co-annihilation, and extends to $m_{3/2} >60$ TeV.  In both frames, to
the left of the stau LSP region we have a portion of the plane where the
relic density is reduced to {\it below} that in (\ref{eq:wmap}). In this
region $M_1,\ M_2$ and $\mu$ are comparable and the LSP is higgsino-like
but has significant bino and wino components. This would facilitate the
observation of direct and indirect LSP signals, but the signal size
would depend on what fraction of DM density is composed of neutralinos.  
In both frames there is
a peak structure around $\alpha\sim 6-8$ (frame {\it a}) or $\alpha\sim
10$ (frame {\it b}) where $2m_{\tz_1}\sim m_A$ and neutralino
annihilation can occur via the $A$-funnel. For larger values of
$\alpha$, very low values of $m_{3/2}$ are required to obtain
consistency with (\ref{eq:wmap}), which can then only occur via neutralino
annihilation through low mass $t$-channel sfermion exchange (bulk
annihilation). 

Turning to negative values of $\alpha$, in frame {\it a}) we see a
region of red points close to $\alpha \sim -1.5$. Near the boundary
between the red and blue points, the neutralino relic density saturates
the observed CDM relic density via BWCA, while in bulk of the region
with the red points, the annihilation is too rapid and leads to a
smaller density of relic neutralinos. As for the model with zero modular
weights, we do not have a corresponding region in frame {\it b}).  

\FIGURE[htb]{
\epsfig{file=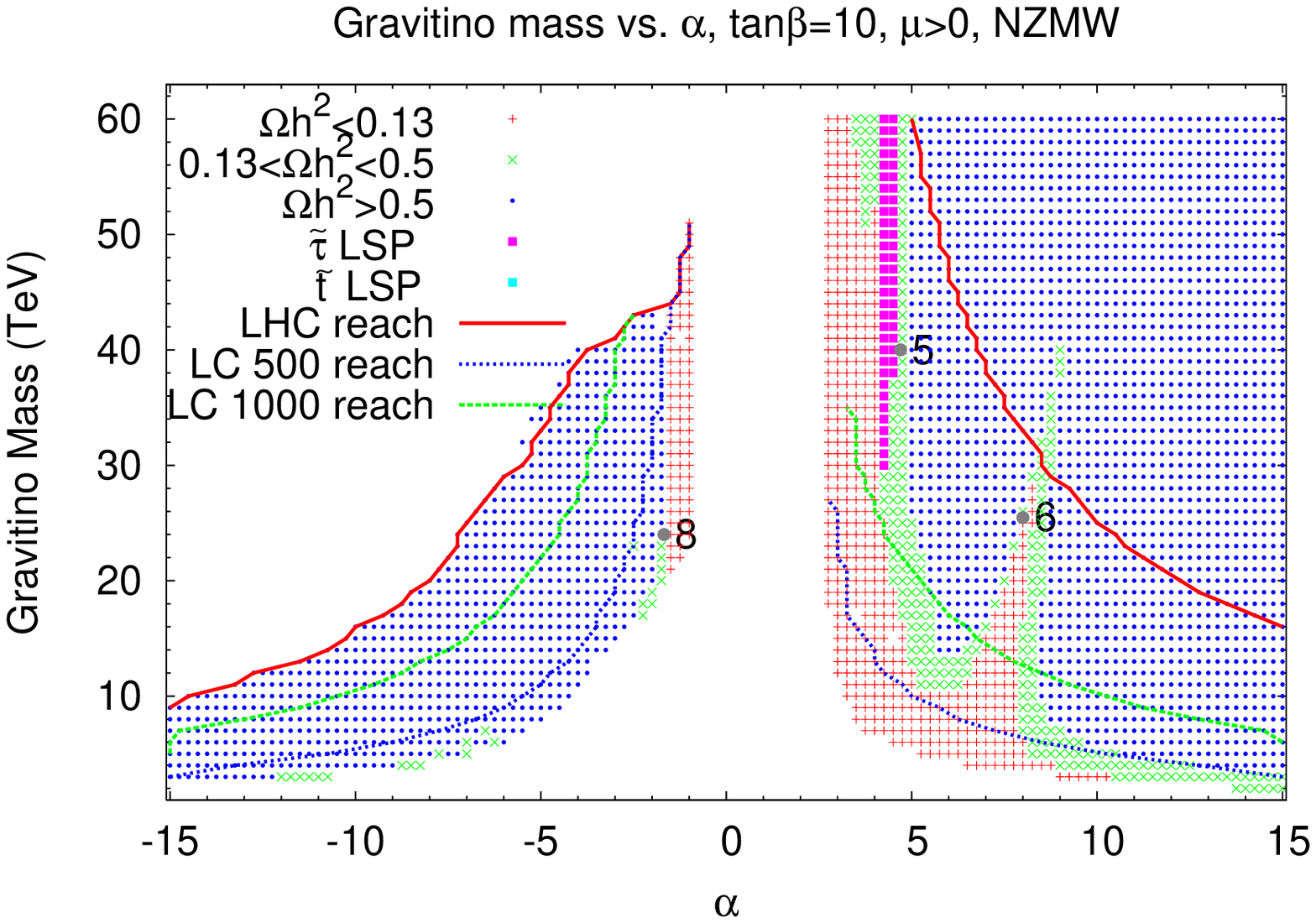,width=12cm} 
\epsfig{file=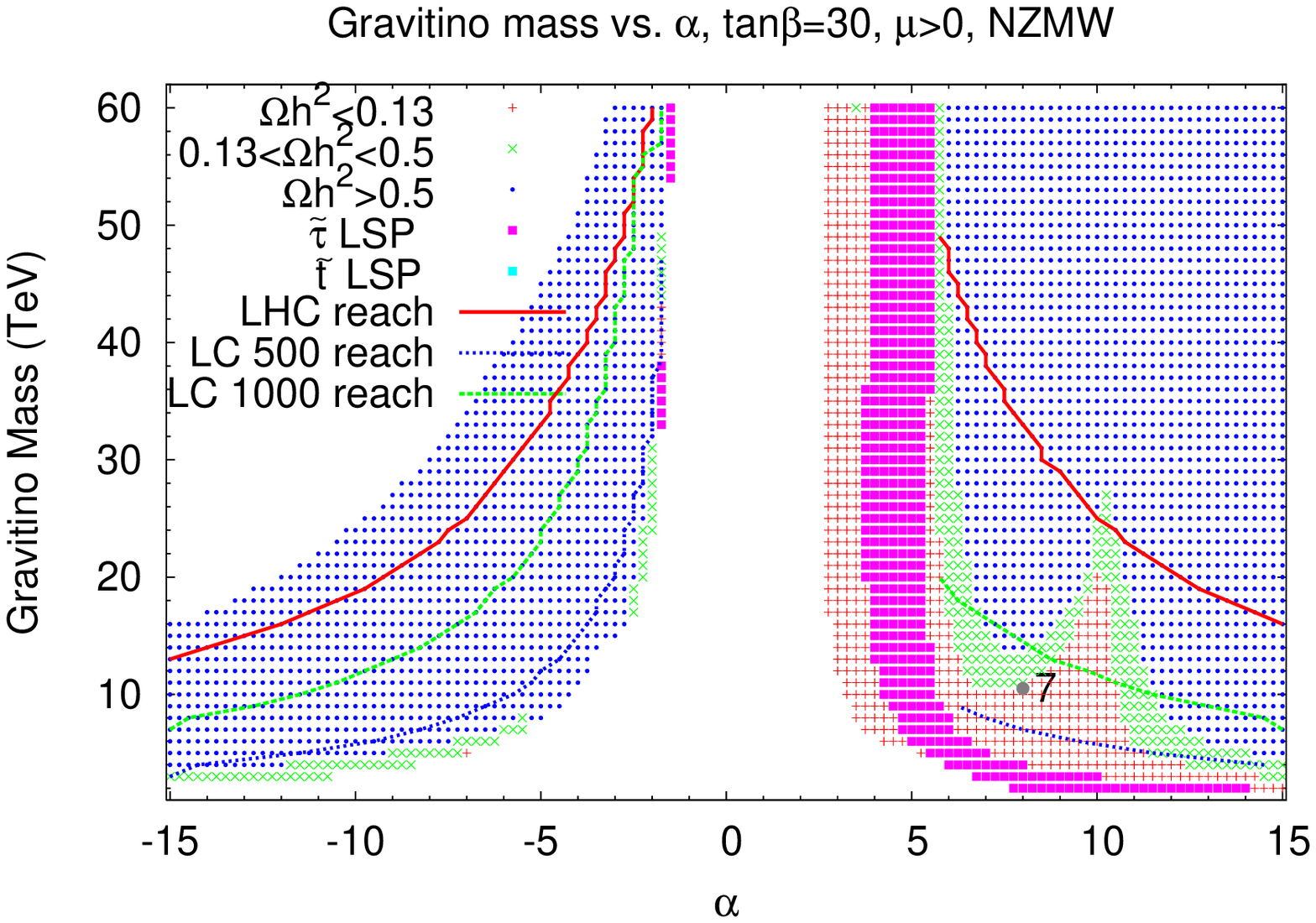,width=12cm} 
\caption{\label{fig:nzmw_plane}
Regions of the $\alpha\ vs.\ m_{3/2}$ plane allowed by theory
constraints, LEP2 searches and neutralino relic density for
{\it a}) $\tan\beta =10$, $\mu >0$ and {\it b}) $\tan\beta =30$, $\mu >0$,
and non-zero modular weights.
We also show the approximate reach of the CERN LHC for
100 fb$^{-1}$ of integrated luminosity, along with the kinematic reach of
a $\sqrt{s}=0.5$ and of a 1 TeV linear $e^+e^-$ collider.}}

We illustrate sample spectra in the MM-AMSB model with non-zero modular
weights with the four points listed in Table \ref{tab:kklt_nzmw}. The
first case, labelled Point~5, has $m_{3/2}=40$ TeV and $\tan\beta =10$,
just as Point~1 of Table \ref{tab:kklt0}, but with $\alpha =4.7$. It is
a stau co-annihilation point. Point~6 is taken at $\alpha =8$,
$m_{3/2}=25.467$~TeV so that it lies in the $A$-annihilation
funnel. Sparticles and non-SM Higgs bosons are relatively heavy for both
these points and give only small contributions to the $BF(b\to s\gamma )$ and
to the anomalous magnetic moment.  Point~7 illustrates a spectrum for
$\tan\beta =30$ with $\alpha =8$, $m_{3/2}=10.5$ TeV, and is located at
the intersection of the stau co-annihilation and $A$-funnel regions.  It
has gluino and squark masses around 1~TeV, while sleptons and the
lighter inos have masses that are accessible at a 1~TeV
$e^+e^-$ collider. The branching fraction for $b\to s\gamma$
is on the low side of its acceptable range and there is a modest SUSY
contribution to the muon $g-2$. We see that the $BF(B_s\to\mu^+\mu^-)$
deviates considerably from its SM value, primarily because of the larger
value of $\tan\beta$.  Finally, we choose Point~8 with a negative value
of $\alpha=-1.64$, with $m_{3/2}=24$~TeV and $\tan\beta=10$ where BWCA
yields to a neutralino relic density in accord with (\ref{eq:wmap}).  We
have checked that for $\alpha$ close to this value, the relic neutralino
density saturates the measured CDM density over the entire range of
$m_{3/2}$ where we have red $+$s in Fig.~\ref{fig:nzmw_plane}: less
negative values of $\alpha$ yield a lighter chargino and the neutralino
relic density becomes too low, so that another source of DM is also
needed in this region. For this case, squark and gluino masses are
similar to those for Point~7, but sleptons, and especially the lighter
charginos and neutralinos are considerably lighter. As for the model
with zero modulus weights, the contribution to muon $g-2$ is small and
negative, while $BF(b\to s\gamma)$ is close to the upper end of its
acceptable range.

%
\begin{table}
\begin{tabular}{lcccc}
\hline
parameter & Point~5 & Point~6 & Point~7 & Point~8 \\
\hline
$\alpha$        & 4.7    & 8  & 8 & -1.64 \\
$m_{3/2} (TeV)$ & 40 & 25.467  & 10.5 & 24 \\
$\tan\beta$     & 10 & 10 & 30 & 10 \\
$\mu$       & 1022.9 & 1425.2 &  625.6 & 816.4 \\
$m_{\tg}$   & 1808.8 & 2304.6 &  1010.6 & 1112.5\\
$m_{\tu_L}$ & 1644.1 & 2158.8 &  948.2 & 1073.9 \\
$m_{\tst_1}$& 1254.5 & 1601.5 &  679.6 & 765.6\\
$m_{\tb_1}$ & 1520.5 & 1971.1 &  832.0 & 953.4\\
$m_{\te_L}$ & 985.7 & 1149.0 &  479.9 & 308.7\\
$m_{\te_R}$ & 886.3 & 993.0 &  412.2 & 199.0\\
$m_{\ttau_1}$ & 879.8 & 982.9 &  352.2 & 185.2\\
$m_{\tw_1}$ & 990.6 & 1110.5 &  441.1 & 135.0\\
$m_{\tz_2}$ & 990.6 & 1108.9 &  440.6 & 134.4\\ 
$m_{\tz_1}$ & 869.1 & 791.2  &  315.0 & 114.1\\ 
$m_A$       & 1135.2 & 1581.2 &  595.8 & 829.4\\
$m_h$       & 119.7 & 120.8 &  116.7 & 114.9\\
$\Omega_{\tz_1}h^2$& 0.12 & 0.12 &  0.11 & 0.10\\
$BF(b\to s\gamma)$ & $3.3\times 10^{-4}$ & $3.4\times 10^{-4}$ &
$2.4\times 10^{-4}$ & $4.1\times 10^{-4}$ \\
$BF(B_s\to\mu^+\mu^-)$ & $3.9\times 10^{-9}$ &$3.9\times 10^{-9}$ &
$6.3\times 10^{-9}$ &$3.8\times 10^{-9}$ \\
$\Delta a_\mu    $ & $1.3 \times  10^{-10}$ & $9.4 \times  10^{-11}$ & 
$16.2\times 10^{-10}$ & $-1.0\times 10^{-10}$ \\ 
$\sigma_{sc} (\tz_1p )$ & $4.4\times 10^{-9}\ {\rm pb}$ & $1.3\times 10^{-10}\ {\rm pb}$ & $2.1\times 10^{-9}\ {\rm pb}$ & $7.8\times 10^{-11}\ {\rm pb}$\\
        $\Phi^{\mu}(km^{-2}yr^{-1})$ & $1.43$ & $0.03$ & $0.45$ & $0.002$ \\
        $\Phi^{\gamma}(cm^{-2}s^{-1})$ & $1.6 \times 10^{-9}\atop (8.3\times 10^{-14})$ & $5.2\times 10^{-7}\atop (2.6\times 10^{-11})$ & 
$2.5 \times 10^{-7}\atop (1.2\times 10^{-11})$ & $9.2\times 10^{-10}\atop (4.7\times 10^{-14})$ \\
        $\Phi^{e^+}(GeV^{-1}cm^{-2}s^{-1}sr^{-1})$ & $4.9\times 10^{-11}\atop (9.6\times 10^{-12})$ & $1.8 \times 10^{-8}\atop (3.7\times 10^{-9})$ 
& $3.7 \times 10^{-9}\atop (8.7\times 10^{-10})$ & $8.3\times 10^{-11}\atop (1.9\times 10^{-11})$ \\
        $\Phi^{\bar p}(GeV^{-1}cm^{-2}s^{-1}sr^{-1})$ & $2.4 \times 10^{-10}\atop (1.4\times 10^{-11})$ & $7.7 \times 10^{-8}\atop (4.4\times 10^{-9})$ 
& $3.2 \times 10^{-8}\atop (1.8\times 10^{-9})$ & $1.0\times 10^{-11}\atop (5.8\times 10^{-13})$ \\
        $\Phi^{\bar {D}}(GeV^{-1}cm^{-2}s^{-1}sr^{-1})$ & $5.0\times 10^{-14}\atop (4.2\times 10^{-15})$ & $2.4 \times 10^{-11}\atop (2.1\times 10^{-12})$ & $1.6 \times 10^{-11}\atop (1.4\times 10^{-12})$ & $4.2\times 10^{-14}\atop (3.5\times 10^{-15})$ \\
\hline
\end{tabular}
\caption{Masses and parameters in~GeV units
for case studies 5--8 within the NZMW framework with
$n_{H_u}=n_{H_d}=1$, and $n_{\rm matter}=1/2$. Also shown are
predictions for low energy observables along with 
cross sections and fluxes relevant to direct and indirect searches
for dark matter. In all cases,
we take $m_t=175$ GeV and $\mu >0$.
The halo annihilation rates use the Adiabatically Contracted N03 Halo model, while the values in
parenthesis use the Burkert halo profile.
}
\label{tab:kklt_nzmw}
\end{table}

Before turning to the discussion of collider signals, we note that by
choosing different values for the modular weights for matter and Higgs
supermultiplets we can obtain the so-called non-universal Higgs mass
(NUHM) model\cite{nuhm}, 
where the GUT scale values of $m_{H_u}^2$ and $m_{H_d}^2$
are split off from the masses of the matter scalars. 
Moreover, if the modular
weights for the Higgs multiplets are chosen to be the same but larger
than those for matter multiplets, the (common) GUT scale Higgs boson
mass squared parameter, $m_{\phi}^2$, is (much) larger than the
corresponding matter parameters, and we get what has been referred to as the
NUHM1 model with large $m_{\phi}^2/m_0^2$.\footnote{More precisely, it
is the NUHM1 model except for the splitting between $m_{H_u}^2$ and
$m_{H_d}^2$ due to the AMSB contribution.} It is then possible for 
$|\mu|$ to become
small enough so that the neutralino is MHDM\cite{nuhm}.
Motivated by these considerations, 
 we examined the case, $n_{H_u}=n_{H_d}=0$, $n_{\rm matter}=1$.
In Fig.~\ref{fig:scan01}, we show the results of our scan of the
$\alpha-m_{3/2}$ parameter plane for this choice of modular weights,
with $\tan\beta =10$. We
see that most of the region of large positive $\alpha$ where we may have
expected to obtain MHDM is excluded because the stau becomes lighter
than $\tz_1$. This situation may be different if we take $n_{\rm
matter}=1/2$, but then the ratio $m_{\phi}^2/m_0^2$ is also reduced; we
have not examined this possibility. For negative values of $\alpha$, we
see that there is a viable region with red points near $\alpha \simeq -2$,
just a bit below where we had obtained the BWCA solution in the previous
figure. We have checked that agreement with (\ref{eq:wmap}) is obtained
via stau co-annihilation, and that the difference 
$|M_2| - |M_1|$ is indeed too large
for BWCA. Consistency with the relic density constraint also obtains for
very small values of $m_{3/2}$ via bulk annihilation: these points,
however, have a rather large negative SUSY contribution to $(g-2)_\mu$,
and so are strongly disfavored. Finally, we
note that unlike previous cases, for this choice of modular weights a
TeV LC will be able to probe ranges of parameters beyond the LHC. This
is primarily because $\ttau_1$, and to a smaller extent, also $\te_R$
and $\tmu_R$ are accessible even for $m_{\tg} \simeq m_{\tq} \agt
3$~TeV.

\FIGURE[htb]{
\epsfig{file=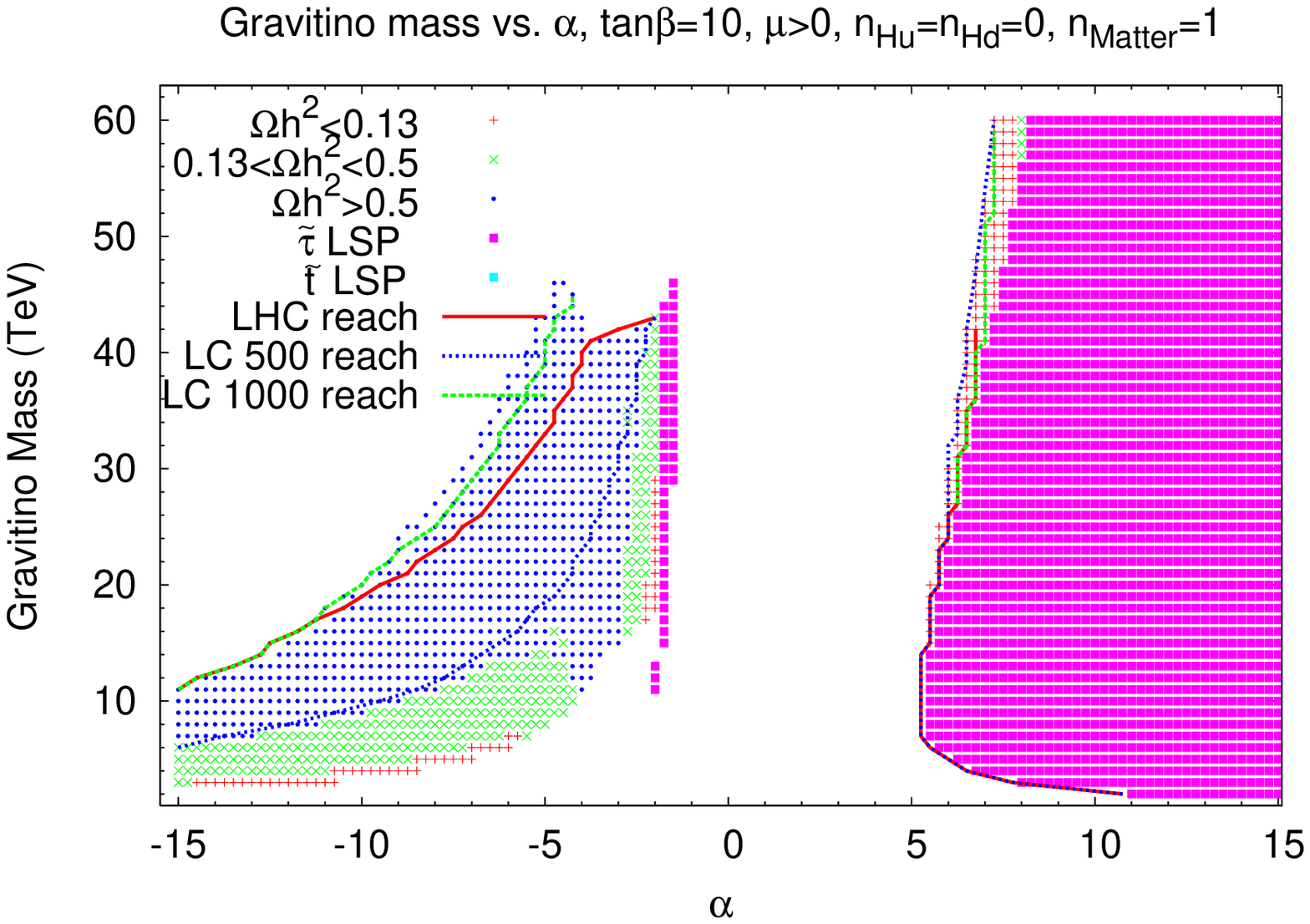,width=12cm} 
\label{fig:scan01}
\caption{
Regions of the $\alpha\ vs.\ m_{3/2}$ plane allowed by theory
constraints, LEP2 searches and neutralino relic density for
$\tan\beta =10$, $\mu >0$ and the choice $n_{H_u}=n_{H_d}=0$ and 
$n_{\rm matter}=1$ for the modular weights.
We also show the approximate reach of the CERN LHC for
100 fb$^{-1}$ of integrated luminosity, along with the kinematic reach of
a $\sqrt{s}=0.5$ and 1 TeV linear $e^+e^-$ collider.}}

\subsection{Prospects for collider and dark matter search experiments}

As in the case of the MM-AMSB framework with zero modular weights, the
NZMW model generally produces SUSY spectra with $m_{\tq}\sim
m_{\tg}$. This is because unless $|\alpha|$ is very small, the GUT scale
gluino mass parameters are comparable to (or larger than) the
corresponding squark parameters, so that bulk of the physical squark and
gluino masses come from the renormalization group evolution to the weak
scale. This is analogous to the more familiar situation in mSUGRA when
$m_0 \alt m_{1/2}$. Since, as we discussed in the last section, the
reach of the LHC extends out to $m_{\tg} \sim m_{\tq} \simeq 3$~TeV for
this case, we expect a similar reach for the NZMW model, assuming an
integrated luminosity of 100~fb$^{-1}$.
This reach is shown by the red contours in Fig. \ref{fig:nzmw_plane},
which generally track the $m_{\tg}\sim 3$ TeV contour, except when they
are on the edge of the allowed parameter space, as for instance for
$\alpha < 0$ in frame {\it a}).  It is noteworthy that the LHC reach
generally encompasses the entire $A$-annihilation funnel. In fact, only
a small stau co-annihilation region with $m_{3/2}> 60$ TeV and $\alpha
\sim 5.5$ at low $\tan\beta$ can escape LHC detection. The LHC with 100
fb$^{-1}$ can cover all the relic-density-allowed parameter space for
$\tan\beta =30$. The reach of a linear $e^+e^-$ collider is mainly
determined by the kinematic reaches for chargino and $\ttau_1$ pair
production. A 500 GeV collider covers only a small portion of the
allowed parameter plane, and even at a 1~TeV LC considerable portions of
the stau co-annihilation region will not be covered. It is, however,
interesting that the entire BWCA region in frame {\it a}) can be probed
at a LC, since it is only via LC experiments that we will have any
chance of directly probing the small mass difference between the
chargino and the LSP.

Points~5 and 6 should be observable via multijet+ multilepton $+\eslt$
signatures at the LHC, although their spectra, other than the light
Higgs, will be inaccessible to either of the LC options. In both cases,
gluinos decay a significant fraction of the time to third generation
squarks, so that LHC SUSY events should be rich in $b$-jets.  It may
also be possible to access the very heavy $\tw_1$ and $\tz_2$ (and for
Point~5, also $\tz_4$ and $\tw_2$) via cascade decays of
$\tq_L$. Point~7 should easily be visible at LHC. Once again, gluinos
decay preferentially to the third generation. Decays of these and of the
light squarks yield $\tw_1$ and $\tz_2$ at large rates. The latter
mostly decays via $\tz_2\to \ttau_1\tau$ and $\tz_2\to h\tz_1$ so that
construction of mass edges may be difficult: see, however Ref.\cite{kamon}. 
Although sparticles are essentially inaccessible at a
500~GeV LC, a 1~TeV collider will provide access to sleptons and
light chargino/neutralino pairs, though some of the rates may be
suppressed by phase space. Finally, at Point~8, there should again be a
plethora of signals at the LHC. Sparticle decay patterns are
qualitatively similar to those for Point~7, but the big difference is
the approximate equality: $m_{\tz_1}\simeq m_{\tw_1} \simeq
m_{\tz_2}$. As a result, the decay products of $\tw_1$ and $\tz_2$ will
be relatively soft at the LHC. In this case, though, $\ttau_1$, $\te_R$
and the lighter charginos and neutralinos should be accessible at even a
500~GeV LC, whereas {\it all} sleptons are accessible at 1~TeV. As
mentioned above, LC experiments may be crucial if the BWCA mechanism is
what reduces the relic density. 

Turning to dark matter searches in the NZMW scenario, we see that none
of the Points 5--8 will lead to an observable signal at CDMS2, though
Points 5 and 7 may be observable at a stage 3 detector such as a 1~ton Xenon
facility. The sensitivity of IceCube is too low for all these points.
Prospects for detection of gamma rays or anti-particles from neutralino
annihilation appear to be better in the NZMW case relative to the case
of zero modular weights. For Point~5 (relative to Point~1 which has
a quite similar spectrum) this is because of the increased higgsino
component in the LSP. Point~6  (7) is in (on the edge of) the
$A$-funnel, so that an increased rate should not be a surprise. Assuming
the same sensitivity as in the last section, we see that there should be
a gamma ray signal in GLAST for all the points using the N03 halo profile, 
while none of the points are observable using the Burkert profile. 
Pamela should have
observable positron and anti-proton signals for Points~6 and
7 (Point 6) in the N03 profile (Burkert profile). 
Points~6 and 7 also give a detectable anti-deuteron signal 
for both halo profiles. 


\section{Summary and Conclusions}
\label{sec:conclude}

The illustration by Kachru {\it et al.}\cite{kklt} 
that compactifications with fluxes
in extra spatial dimensions in type IIB string models can stabilize the
moduli and give rise to a de Sitter vacuum for the Universe has spurred
several recent studies of the structure of the soft SUSY breaking terms
in these scenarios. The structure of these terms depend on certain
integers or half integers, the so-called modular weights of the MSSM
superfields, that characterize their location in the extra
dimensions. Phenomenologically, the most important feature of this
scenario is that the SSB parameters can obtain comparable contributions
from the mediation of SUSY breaking by moduli fields and the so-called
anomaly mediation of SUSY breaking, in contrast to previously studied
models where anomaly-mediated SUSY breaking effects were considered to
be negligible unless mediation by moduli is, for one reason or other,
essentially negligible. The relative strength of modulus and
anomaly-mediated contributions is controlled by a phenomenological
parameter $\alpha$ which can assume any (positive or negative) real
number. The value of $\alpha$ (and many other parameters) will
be fixed if it ever becomes possible to explicitly construct a realistic
vacuum starting from string theory. Until this time, we have to use
this construction solely as a motivation for the examination of the
phenomenology of MM-AMSB models.

Flavor changing neutral current constraints suggest that there should
not be large mass splittings between sparticles with
the same gauge quantum numbers. This can be ensured by choosing 
common modular weights for super-multiplets with the same gauge quantum
numbers. This consideration, by itself, does not constrain the modular
weights of different MSSM multiplets, but of course, if we want to embed
these in a GUT, we should choose a common modular weight for all
particles in the same  GUT multiplet. Viewed differently, these should all 
be at the same location in the extra dimension. Throughout the paper, we
assume that all gauginos also reside at a common location, and so
receive a common mass from modulus mediation. 

We have begun our phenomenological study assuming, for simplicity, that
the modular weights are {\it all} zero. This gives a universal
modulus-mediated contribution to all GUT scale scalar mass parameters,
to the trilinear $A$-parameters and to the gaugino masses. Moreover, the
GUT scale values of these contributions are in the ratio,
$m_0:m_{1/2}:A_0= 1:1:-3$. Of course, the usual anomaly-mediated
contribution has also to be included, and for scalar mass parameters,
there is a mixed modulus-anomaly-mediated contribution also. The GUT
scale values of these SSB parameters are given by
(\ref{eq:M})--(\ref{eq:A}). The framework is more general than the
well-studied mSUGRA or AMSB frameworks in that $\alpha$ allows us to
control the relative strengths of the respective contributions, 
but is more constrained in that the universal contributions to all SSB
parameters are in a fixed ratio. Indeed, once the modular weights are
fixed, the parameter space is {\it smaller} than that of the mSUGRA
model. 

Turning to the phenomenology of the model with zero modular weights, we
find that the spectrum is characterized by a relatively light $t$-squark
or tau slepton. This is because of the large value of the
$A$-parameter. Indeed, for positive values of $\alpha$, consistency with
the observed relic density is usually obtained only when the LSP, which
is mostly bino-like, can co-annihilate with either $\tst_1$ or
$\ttau_1$: Higgs funnel annihilation is possible for a limited range of
parameters. Because the ratio $M_1:M_2$ can be adjusted by an
appropriate choice of $\alpha$, it may appear that it should be possible
to adjust it to obtain MWDM (by setting 
$M_1({\rm weak})\simeq M_2({\rm  weak})$) or BWCA (by setting 
$M_1({\rm weak})\simeq -M_2({\rm   weak})$). 
We found, however, that MWDM is not possible because the
  $\tst_1$ becomes the LSP for the required 
value of $\alpha$. It is, however, possible to obtain agreement with the
relic density constraint via BWCA by choosing $\alpha\sim
-1.7$. Experiments at the LHC will essentially probe the entire range of
parameters of the MM-AMSB model with zero modular weights for parameters
consistent with the determination of the DM relic density. However,
because the LSP is mostly a bino that co-annihilates with the stop or
the stau, signals for direct and indirect detection of neutralino DM are
usually small. 

We also presented results for one case of non-zero modular weights, with
$n_{H_u}=n_{H_d}=1$ and $n_{\rm matter}={1\over 2}$. In this case, the top
squark mass is much heavier than in the ZMW case. However, WMAP allowed
regions can still be found via either stau co-annihilation, $A$-funnel
annihilation, mixed higgsino/wino DM or via BWCA.  The CERN LHC can
cover almost all of the interesting parameter space.  A combination of
LHC and LC measurements may allow the SSB parameters to be extracted:
their extrapolation to higher scales via RGE should exhibit ``mirage
unification'', as do the ZMW solutions. Prospects for DM detection are
somewhat better for this choice of modular weights, primarily because
$A$-funnel annihilation is possible for some ranges of model
parameters. 

In summary, we have explored the phenomenology of a novel MM-AMSB framework
where MSSM SSB parameters receive comparable modulus- and
anomaly-mediated SUSY breaking contributions. The framework leads to
unusual sparticle mass patterns, and can naturally accommodate the BWCA
scenario. Experiments at the LHC will be able to explore almost all parameter
regions consistent with the measured CDM density, while only limited
ranges of the parameter space will be available to LC
experiments. Fortunately, experiments at a LC will be able to perform
detailed studies of charginos and neutralinos  if nature chooses the
BWCA mechanism to make the relic density in accord with
observation. These studies are difficult at the LHC on account of the
small mass gaps between $\tz_1, \ \tz_2$ and $\tw_1$. 
In the case that the 
soft terms are able to be extracted via a combination of LHC and LC
measurements, then the phenomenon of ``mirage unification''
should be evident, especially for the gaugino masses.

\section*{Acknowledgments}
 
We thank K. Choi, A. Falkowski and Y. Mambrini for helpful correspondence.
This research was supported in part by the U.S. Department of Energy
grants DE-FG02-97ER41022 and DE-FG03-94ER40833.

\section*{NOTE ADDED} 
The phenomenology of the model with NZMW was also examined by Kitano and
Nomura \cite{KN}. In their analysis, they fix the mirage unification
scale to be $\sim$~TeV which, in turn, implies $\alpha \simeq 3$,
corresponding to the small $\mu$ solutions that we find in
Fig.~\ref{fig:nzmw_plane}. However, while we obtain $m_A$ and $\mu$ via
two loop renormalization group evolution using the boundary values of
$m_{H_u}^2$ and $m_{H_d}^2$ as given in (\ref{eq:m2}), Kitano and Nomura,
who essentially perform a one-loop analysis, argue that two loop
renormalization group effects would generate non-zero values of the
Higgs boson squared mass parameters at the mirage unification scale, and
so trade these for $\mu$ and $m_A$ which they treat as free parameters
that they vary in the range $|\mu| < 190$~GeV and $m_A < 300$~GeV, where
the range follows from the requirement that their naturalness parameter
$\Delta < 0.2$. This same requirement bounds the gravitino mass scale
giving them a light sparticle spectrum, corresponding to the small
$\alpha > 0$ slice of the plane of Fig.~\ref{fig:nzmw_plane}. We see,
however, from Fig.~\ref{fig:sm_evol_nz} that while two loop effects do
indeed make $m_{H_u}^2$ and $m_{H_d}^2$ non-zero (and negative) at the
mirage unfication scale, they evolve to a specific value, so that the
values of $|\mu|$ and $m_A$ are completely calculable within this
framework.  Finally, we note that the origin of mirage unification for
the NZMW model (in fact in all models where $n_i+n_j+n_k=2$, where $i$,
$j$ and $k$ are the fields that enter the Yukawa couplings and trilinear
soft terms; {\it e.g.} the model illustrated in Fig.~\ref{fig:scan01})
at the one loop level can be understood from the formulae given by Choi
{\it et al.} \cite{choi3}, as well as from the analysis in the Appendix
of the second paper in Ref.\cite{KN}. We thank R. Kitano and Y.~Nomura
for bringing their work on the MM-AMSB model to our attention.

%

\end{document}